\documentclass[aps,prd, preprint, onecolumn,superscriptaddress,amssymb,amsmath,nofootinbib]{revtex4-1}
\pdfoutput=1

\usepackage{color}
\usepackage{epsfig}
\usepackage{graphicx}
\usepackage{float}
\usepackage{epstopdf}
\usepackage{hyperref} 
\usepackage[usenames,dvipsnames]{xcolor}
\usepackage{slashed}
\usepackage{braket}
\usepackage{mathtools}
\begin{document}

\preprint{EFI-20-16}

\title{The Scale of Superpartner Masses and Electroweakino Searches \\at the High-Luminosity LHC}

\author{Jia Liu}
\email{jialiu@pku.edu.cn}
\affiliation{School of Physics and State Key Laboratory of Nuclear Physics and Technology, Peking University, Beijing 100871, China}
\affiliation{Center for High Energy Physics, Peking University, Beijing 100871, China}
\affiliation{\mbox{Physics Department and Enrico Fermi Institute, University of Chicago, Chicago, IL 60637}}

\author{Navin McGinnis}
\email{nmmcginn@indiana.edu}
\affiliation{Physics Department, Indiana University, Bloomington, IN 47405, USA}
\affiliation{High Energy Physics Division, Argonne National Laboratory, Argonne, IL, 60439}

\author{Carlos E.M. Wagner}
\email{cwagner@anl.gov}
\affiliation{\mbox{Physics Department and Enrico Fermi Institute, University of Chicago, Chicago, IL 60637}}
\affiliation{High Energy Physics Division, Argonne National Laboratory, Argonne, IL, 60439}
\affiliation{\mbox{Kavli Institute for Cosmological Physics, University of Chicago, Chicago, IL, 60637}}

\author{Xiao-Ping Wang}
\email{hcwangxiaoping@buaa.edu.cn}
\affiliation{\mbox{School of Physics, Beihang University, Beijing 100083, China}}
\affiliation{High Energy Physics Division, Argonne National Laboratory, Argonne, IL, 60439}


\date{\today}

\begin{abstract}
Searches for weakly interacting particles is one of the main goals of the high luminosity
LHC run. In this work we study the well motivated cases of electroweakinos with mostly
Wino and Bino components. We show the relevance of squark induced t-channel production in defining the production cross section and hence the LHC reach.
 Moreover, a realistic evaluation of the decay branching ratios show a strong dependence
 on the sign of $\mu$ and, for negative values of $\mu$, on the relative size of the ratio of 
 $\mu$ to the gaugino masses compared with $\tan\beta$.  Overall, unless it is kinematically
 suppressed, or specific conditions are fulfilled, the Higgs decay channel is the most significant 
 one, and the trilepton channel becomes subdominant with respect to final states including
 bottom quarks.  Although the properties are different than in the Higgsino-Bino case,
 also in this case the discovery reach extends to mass values that are significantly larger than the 
 ones  probed at current luminosities, leading to a strong motivation for the search for 
 electroweakinos in the high luminosity LHC run.
\end{abstract}


\maketitle

\tableofcontents




\section{Introduction}

Searches for new particles at the LHC have so far provided no evidence of the existence of new physics 
at the TeV scale. These searches are quite sensitive to strongly interacting particles and have excluded
the presence of vector like quarks or, in the case of supersymmetry, squarks and gluinos, for masses beyond 1~TeV. For examples of recent searches see Refs.~\cite{Aaboud:2017vwy, Sirunyan:2018vjp}.   It is, however, premature to announce the absence of new physics at the 
electroweak scale due to these observations.  On one hand, these searches have been mostly interpreted within
simplified models with simple decay channels designed to maximize the observability of new particles and
hence the bounds may be relaxed in the case of more complicated decay channels. More 
importantly, the searches become mostly insensitive to weakly interacting particles for which
the production cross sections become much weaker than the strongly interacting particle ones.

Weakly interacting particles are naturally involved in one of the main hints for physics at the weak
scale, namely Dark Matter~\cite{Jungman:1995df,Bertone:2004pz}.  The Dark Matter particle appears naturally as part of the weakly
interacting sector of extensions of the Standard Model, in a similar way to the appearance of
the neutrino in the Standard Model (SM) of particle physics.  For heavy particles with masses
of the order of the weak scale, the Dark Matter particle is identified with the lightest of these
new particles and the stability of these neutral and weakly interacting particles demand the 
presence of a symmetry, usually discrete (such as R-parity in the MSSM), that forbids the decay of this particles into SM ones.   
Production of these beyond the SM particles leads to decays into Higgs and
weak gauge bosons and the Dark Matter particle which is observed as missing energy.

A particularly well motivated electroweak sector that has been studied in quite detail both
theoretically as well as experimentally is the one implied by low-energy supersymmetric extensions,
and in particular the one associated to the Minimal Supersymmetry Extension of the SM
(MSSM)  \cite{Fayet:1976et, Fayet:1977yc, Gunion:1987kg, Haber:1984rc,Martin:1997ns,
Sirunyan:2017zss, Sirunyan:2017lae, Sirunyan:2018ubx, Sirunyan:2018iwl, Aaboud:2018jiw, Aaboud:2018sua, Aaboud:2018ngk, Aad:2019vvf, ATLAS-CONF-2019-008, ATLAS-CONF-2019-020, Aad:2020qnn}. In this case, the electroweak sector consists of two Higgs doublets and their
corresponding superpartners (Higgsinos) as well as the superpartners of the weak and
hypercharge gauge bosons (Winos and Bino, respectively).  The couplings of these
particles to the gauge bosons and the Higgs bosons are dictated by the invariance 
under gauge and supersymmetry transformations, the latter being violated only softly 
by dimensionful parameters.   These mass parameters include the Wino $M_2$ and Bino
$M_1$ masses, as well as the Higgsino  mass parameter $\mu$.  The Higgs sector
is characterized by the mass of the CP-odd Higgs $m_A$ and $\tan\beta$, the ratio
of the Higgs vacuum expectation values.   Due to the supersymmetry relations, the
mass of the colored particles also play a role in determining the lightest Higgs mass
and also contribute to the t-channel production cross section for gaugino-like particles (the Higginos
couple very weakly to the first and second generation quarks.)~\cite{Baer:1993tr,Baer:1994nr,Lopez:1994dm,Baer:1995nq,Baer:1995bu,
Yu:2014mda}.

In a previous article \cite{Liu:2020muv}, we studied the search for particles in the Higgsino-Bino sector
of this model, assuming that the Wino mass is of order of a few TeV and a decoupled sfermion spectrum. We demonstrated
the complementarity of the production of electroweakinos via the heavy Higgs
bosons with the ones induced by the direct production of these particles via 
gauge bosons and, due to the smaller production cross sections, 
we  showed that the regions probed at present  are far weaker than
the ones that are usually displayed experimentally for the Wino  case.
Moreover, we showed that the discovery reach of the high luminosity 
LHC go far beyond the current probed region.  The final states including gauge
and Higgs bosons played a similarly relevant role in this analysis.

In this work, we extend this analysis to the Wino case, which differs from the
Higgsino case in several relevant aspects. On one hand the production cross
section has a relevant dependence on the masses of the first and second generation
squarks.  On the other hand, the branching ratios of the decay of the neutral
Winos into Higgs and $Z$ final states depend  on the sign of $\mu$,
the Higgs decay being in general dominant for positive $\mu$, and also
for negative values of $\mu$ unless one is in the proximity of a so-called
blind spot solution, that occurs when the ratio of $|\mu|$ to the average gaugino
masses is of order $\tan\beta/2$.  This implies a more complex (and weaker)
reach for Winos to the one that is usually shown in experimental searches, 
that rely on large branching ratios and very heavy squark masses. 

This work is organized as follows. In section \ref{sec:mass}, we briefly review the mass eigenstates
and mixing for the Wino case and calculate its decay branching ratio to $Z$ and SM Higgs.
In section \ref{sec:prod-br}, we point out the squark mass dependence for the production 
cross section and show the parametric dependence of the Wino decay branching ratios on $\mu$ and $\tan\beta$.
In section \ref{sec:reach}, we show the resultant, current bounds and future reach of the electroweakino 
searches. We reserve section \ref{sec:conclusion}, for our conclusions.

\section{Mass eigenstates and couplings to $Z$ and SM Higgs}
\label{sec:mass}

The mass eigenstates and decays modes of all electroweakinos are determined by only four parameters, the Wino and Bino masses $M_{2}$ and $M_{1}$, the Higgsino mass $\mu$, and the ratio of vacuum expectation values of the Higgs doublets $\tan\beta$. The resulting mass matrices for the neutralinos and charginos in terms of these parameters are given by 

\begin{flalign}
M_{N}=&\begin{pmatrix}
M_{1} & 0 & -c_{\beta}s_{W}m_{Z} & s_{\beta}s_{W}m_{Z}\\
0 & M_{2} &c_{\beta}c_{W}m_{Z} & -s_{\beta}c_{W}m_{Z}\\
-c_{\beta}s_{W}m_{Z} & c_{\beta}c_{W}m_{Z} & 0 & -\mu\\
s_{\beta}s_{W}m_{Z} & - s_{\beta}c_{W}m_{Z} & -\mu & 0
\end{pmatrix},\\
&M_{C}=\begin{pmatrix}
M_{2} & \sqrt{2}s_{\beta}m_{W}\\
 \sqrt{2}c_{\beta}m_{W} & \mu
\end{pmatrix},
\end{flalign}
where $c_{W}=\cos\theta_{W}$, $s_{W}=\sin\theta_{W}$, $m_{Z}$ and $m_{W}$ are the $Z$ and $W$ gauge boson masses, and $\theta_{W}$ is the weak-mixing angle. For further details of the couplings and mass matrices of neutralinos and charginos see, for instance, Ref.~\cite{Martin:1997ns}.

The neutralino mass eigenstates are given after diagonlization $Z_N^T M_{N} Z_{N} = \tilde{m}_{\chi}$ where $\tilde{m}_{\chi}={\rm diag}(m_{\chi_1^0},m_{\chi_2^0},m_{\chi_3^0},m_{\chi_4^0})$ and the mixing matrix, $Z_{N}$, encodes the admixtures of the gauge eigenstates in the neutralinos. In general, the particular form of $Z_{N}$ is not particularly illuminating. However, in this paper we will focus on the case when the Higgsinos are heavy and the low energy spectrum consists of Wino- and Bino-like states. In the limit of $|\mu| \gg M_1, M_2$, we find for the mixing matrix $Z_N$

\begin{align}
\tiny
\left(
\begin{array}{cccc}
1-\frac{m_Z^2s_w^2}{2\mu^2}\left( 1+\frac{m_Z^2c_w^2s^2_{2\beta}}{\left( M_2-M_1\right)^2}\right)&\frac{m_Z^2s_{2w}}{2\mu^2\left( M_2-M_1\right)}\left[ \mu s_{2\beta} + M_2 +\frac{m_Z^2s^2_{2\beta} c_{2w} }{ M_2-M_1}\right] & -\frac{m_Zs_w\left( \sin_\beta +\cos_\beta\right)\left( \mu+M_1\right)}{\sqrt{2}\mu^2} &  \frac{m_Zs_w\left( \cos_\beta -\sin_\beta\right)\left(\mu-M_1 \right)}{\sqrt{2}\mu^2}   \\
\frac{m_Z^2s_{2w}}{2\mu^2\left( M_1-M_2\right)}\left[ \mu s_{2\beta} + M_1 +\frac{m_Z^2s^2_{2\beta} c_{2w} }{ M_2-M_1}\right]   &1-\frac{m_Z^2c^2_w}{2\mu^2}\left(1+\frac{m_Z^2s_w^2s^2_{2\beta}}{2\left( M_1-M_2\right)^2}\right) & \frac{m_Zc_w\left( \sin_\beta +\cos_\beta\right)\left( \mu+M_2\right)}{\sqrt{2}\mu^2}   &  -\frac{m_Zc_w\left( \cos_\beta -\sin_\beta\right)\left(\mu -M_2\right)}{\sqrt{2}\mu^2}    \\
\frac{m_Z  s_w}{\mu^2} \left[\left( \mu s_{\beta} + M_1c_\beta\right)+\frac{m_Z^2c^2_ws_{2\beta} s_\beta }{\left( M_2-M_1\right)}\right]   &-\frac{m_Zc_w }{\mu^2} \left[\mu s_{\beta} + M_2c_\beta+\frac{m_Z^2s_w^2s_{2\beta} s_\beta}{\left( M_1-M_2\right)}\right] & \frac{1}{\sqrt{2}} -\frac{m_Z^2s_\beta\left(s_\beta+c_\beta \right)}{2\sqrt{2}\mu^2} & \frac{1}{\sqrt{2}}+\frac{m_Z^2s_\beta\left(c_\beta-s_\beta \right)}{2\sqrt{2}\mu^2}   \\
-\frac{m_Zs_w}{\mu^2} \left[\mu c_{\beta} + M_1s_\beta+\frac{m_Z^2c^2_ws_{2\beta} c_\beta }{\left( M_2-M_1\right)} \right] &\frac{m_Z c_w}{\mu^2}\left[ \mu c_\beta+M_2s_\beta + +\frac{m_Z^2 s_{2\beta} c_\beta s^2_w}{M_1-M_2}\right]& -\frac{1}{\sqrt{2}}+\frac{m_Z^2c_\beta\left(s_\beta+c_\beta \right)}{2\sqrt{2}\mu^2}  & \frac{1}{\sqrt{2}} +\frac{m_Z^2c_\beta\left(s_\beta-c_\beta \right)}{2\sqrt{2}\mu^2}
\end{array}
\right).
\end{align}

Assuming $M_2 > M_1$, the mass eigenvalues of the NLSP and LSP are then simply approximated by
\begin{align}
m_{\chi^0_1}&=M_1-\frac{m_Z^2}{\mu}s_w^2 \left(\sin_{ 2\beta } +\frac{M_1}{\mu}\right)\simeq M_1 , \nonumber \\
m_{\chi^0_2}&=M_2-\frac{m_Z^2}{\mu}c_w^2 \left(\sin _{2\beta } +\frac{M_2}{\mu}\right)\simeq M_2,
\end{align}
where for large $|\mu|$ we see that $\chi_{2}^{0}$ and $\chi_{1}^{0}$ are almost pure Wino and Bino mixtures respectively.\\
The parameters that determine the mass eigenstates and mixing also determine the couplings of electroweakinos to gauge and Higgs bosons through mixing of D-terms. In particular, the couplings of $\chi_{2}^{0}$ to $h/Z$ and $\chi_{1}^{0}$, $g_{h\chi^0_1\chi^0_2}$ and $g_{Z\chi^0_1\chi^0_2}$,  can be simplified assuming large $\mu$, $|m_{\chi_1^0} - m_{\chi_2^0}| > m_h, m_Z$, and the Higgs alignment condition $\alpha \approx \beta - \pi/2$. 
We find \footnote{Our result in $g_{h\chi^0_1\chi^0_2}$ has a minor difference with Ref.~\cite{Han:2013kza}, which omitted $m_{\chi_1^0}$,  probably  assuming it to be small.}

\begin{align}
g_{h\chi^0_1\chi^0_2} &= - \frac{e m_Z }{ \mu}\left[ s_{2\beta} +\frac{m_{\chi_1^0} + m_{\chi_2^0}}{2\mu}  + \frac{m_Z^2s^2_{2\beta} c_{2w}}{\mu\left(m_{\chi_2^0}-m_{\chi_1^0}\right)} \right], \\
g_{Z\chi^0_1\chi^0_2}&= - \frac{e m_Z^2 }{2 \mu^2} \left[c_{2\beta}+\frac{m_Z^2s_{4\beta}c_{2w}}{2\mu\left( m_{\chi_2^0}-m_{\chi_1^0}\right)} \right].
\end{align}

We see that when $\mu < 0$, the Higgs coupling $g_{h\chi^0_1\chi^0_2}$ has a blind spot when $s_{2\beta} +(m_{\chi_1^0} + m_{\chi_2^0})/(2\mu)  \sim 0$, and thus along this direction the coupling to the Higgs is suppressed and the neutral Wino decays almost exclusively through the Z-channel. Further, for a given pair of neutralino masses the blind spot is determined by the values of $\tan\beta$ and $\mu$. For example, given $m_{\chi_1^{0}} + m_{\chi_2^{0}} = 800$ GeV, the cancellation happens for $\{ \tan\beta, ~ \mu (\text{GeV})\} =\{ 5,  -1040 \}$, $\{ 10,  -2020 \}$ or $\{ 20,  -4010 \}$.

The decay widths of $\chi_2^0 \to \chi_1^0 ~h$ and $\chi_2^0 \to \chi_1^0 ~Z$ are given by
\begin{align}
\Gamma_h & = \frac{g_{h\chi^0_1\chi^0_2}^2}{8\pi} p_h \frac{\left(m_{\chi_1^0} + m_{\chi_2^0} \right)^2 -m_h^2 }{m_{\chi_2^0}^2 }, \\
\Gamma_Z & = \frac{g_{Z\chi^0_1\chi^0_2}^2}{8\pi} p_Z \frac{\left(m_{\chi_1^0} + m_{\chi_2^0} \right)^2 -m_Z^2 }{m_{\chi_2^0}^2 } \times
\frac{\left(m_{\chi_2^0} - m_{\chi_1^0} \right)^2 + 2 m_Z^2 }{m_{Z}^2 } ,
\end{align}
where $p_h$ and $p_Z$ are the momentum of $h$ and $Z$ in the final state. These results are in agreement with Ref.~\cite{Choi:2003fs}. For a review of electroweakino scenarios in the MSSM and corresponding decay modes see Ref.~\cite{Canepa:2020ntc}.\\

\subsection{Comments on the anomalous Magnetic Moment and Dark Matter for large values of $|\mu|$}

The SM prediction for the muon anomalous magnetic moment, $(g-2)_\mu$ differs by about 3.5~standard deviations with respect
to the current experimental value measured at the Brookhaven $g-2$ experiment~\cite{Aoyama:2020ynm,Blum:2018mom,Bennett:2006fi}, 

\begin{equation}
\delta a_\mu^{\rm exp} \simeq \left( 27 \pm 7 \pm 5   \right) \times 10^{-10},
\end{equation} 
where the errors are associated with experimental and theoretical uncertainties. 

The dominant contribution to the muon anomalous magnetic moment in the MSSM~\cite{Ellis:1982by,Moroi:1995yh,Carena:1996qa,Czarnecki:2001pv,Feng:2001tr,Martin:2001st}, comes from the chargino induced diagram.  
This contribution relies on the presence of a relevant Higgsino and Wino component of the light charginos and hence is suppressed when $|\mu|$ becomes sizable. For large values of $|\mu|$, like the ones
analyzed in this article, the neutralino contribution may be also relevant. Contrary to Winos, Binos couple to both left- and right-handed muons and tend to provide the most 
relevant contributions through mixing in the slepton sector, with the parameter $\mu$ providing the necessary left-right slepton mixing.  Since this mixing is proportional to $\tan\beta$, sizable neutralino contributions may be obtained for large 
values of $\mu \tan\beta$ and sleptons that may easily be heavier than the characteristic Wino mass scale discussed in this article.   We verified these properties quantitatively by using the code 
CPsuperH~\cite{Lee:2003nta,Lee:2012wa}. As an explicit example, if one takes the extreme values of $\mu = 10$~TeV
and $\tan\beta = 50$, considered in this article, values of $M_1$ of the order of the weak scale and first and second generation slepton masses of order 700 GeV will be necessary to get the current experimental 
central value for the muon $g-2$.  On the other hand, the third generation sleptons must be heavy in order to avoid problems in the slepton spectrum. 
For smaller values of $\mu$ and $\tan\beta$, the chargino contributions become relevant. For instance, for $\mu = 2$~TeV and $\tan\beta = 50$, $M_2 = 400$~GeV and 
$M_1 =250$~GeV, sleptons with masses of order 450~GeV will lead to the proper $g-2$ contribution, with neutralinos providing about 60 percent of the total contribution.

Dark Matter, on the other hand, may be obtained either by resonant annihilation with scalars or by co-annihilation with the slepton states~\cite{Jungman:1995df,Bertone:2004pz,Drees:1992am}.
For co-annihilation, light sleptons should be close in mass to the Bino states and therefore will lead to additional decays of the the heavier Winos into charged and neutral Wino states, which would be in conflict
with the heavy scalar assumption of the current work. This can only be avoided in the case of light right-handed sleptons, with small mixing with their left-handed partners.  Such small mixing would
suppress the neutralino contribution to $(g-2)_{\mu}$ and is difficult to achieve for large values of $|\mu|$ and $\tan\beta$, but is still possible if one assumes a very large hierarchy between the left- and right-slepton
masses.  Regarding the resonant annihilation via the Higgs states, this may be achieved for moderate values of $\tan\beta$ and values of the Bino mass close to a half of the heavy Higgs masses (the
resonant annihilation contribution via the lightest Higgs (or the $Z$) is highly suppressed for large values of $\mu$~ \cite{Ellis:1999mm,Han:2016qtc, Abdughani:2017dqs, Duan:2017ucw, Carena:2018nlf,Carena:2019pwq, Cao:2019qng} ).  Heavy Higgs
boson annihilation, on the other hand, will be subject to strong LHC constraints, unless $\tan\beta$ is not very large~\cite{Aad:2020zxo,Sirunyan:2018zut}.  For instance, we verified using MicrOmegas~\cite{Belanger:2010pz} that the proper relic density can be obtained for $\tan\beta = 5$, $\mu = 2$~TeV,  and  $M_H \simeq 600$~GeV~\footnote{Masses of the heavy Higgs bosons $\gtrsim 500$ GeV will have a negligable effect on the discussion of electroweakino branching ratios in the following section.} provided $M_1$ is of order 290~GeV, where this value increases to  298~GeV for $\mu = 5$~TeV. The
direct detection cross section tends to be suppressed, an order of magnitude or more below the current bounds, due to the large Higgs and $Z$ coupling suppression induced by the large
values of $|\mu|$.

\section{Production and branching ratios}
\label{sec:prod-br}

\begin{figure}[H]
\centering
	 \includegraphics[scale=0.75]{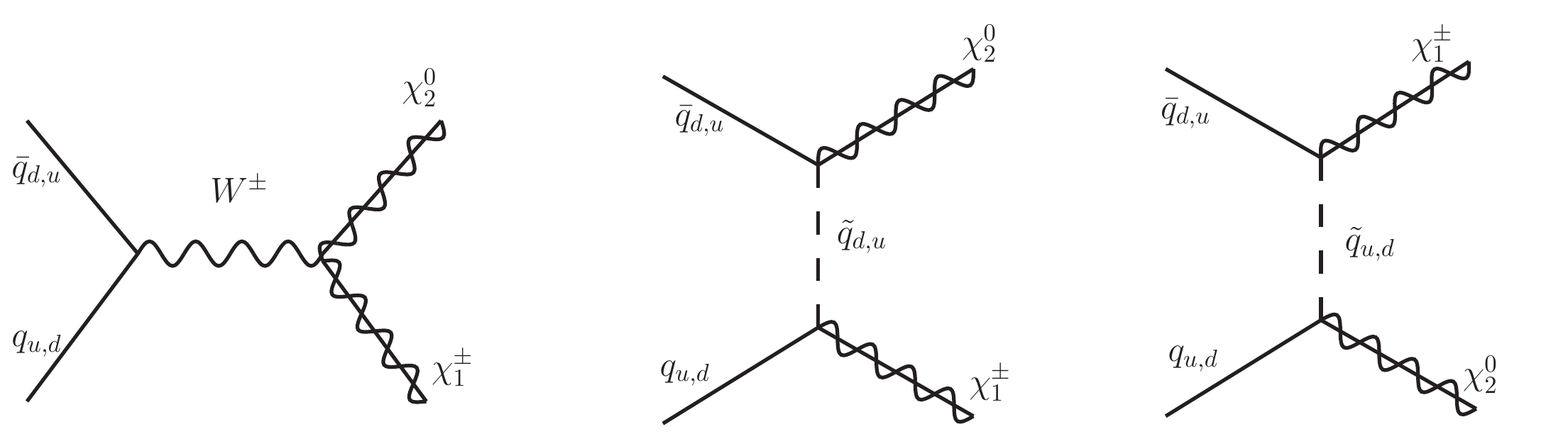} 
\caption{Leading order diagrams contributing to the direct production of electroweakinos at the LHC in the case that the spectrum is Wino-like.}
\label{fig:diagrams}
\end{figure} 

At the LHC, the production of Wino-like electroweakinos, $\chi_{1}^{\pm}$ and $\chi_{2}^{0}$, proceeds mostly through s-channel exchange of a $W$ boson. However, for heavy squarks, the $\chi_{1}^{\pm}$-$\chi_{2}^{0}$ pair is subdominantly produced through t-channel exchange of first- and second- generation squarks~\cite{Baer:1993tr,Baer:1994nr,Lopez:1994dm,Baer:1995nq,Baer:1995bu}, see Fig.~\ref{fig:diagrams}~\footnote{The same is true for other scenarios such as the Higgsino-Bino scenario. However, in such cases the dependence of the couplings to squarks is proportional to their Yukawa couplings and hence negligible.}. Apart from the parametric dependence described in the previous section, the overall production modes of $\chi_{1}^{\pm}$ and $\chi_{2}^{0}$ will also have a dependence on the scale of superpartners, $M_{SUSY}$. The measurement of the Higgs boson mass indicates that stop masses are around $1-10$ TeV in the MSSM~\cite{Draper:2013oza,Bagnaschi:2014rsa,Vega:2015fna,Lee:2015uza,Bahl:2017aev}. Further, exclusion of squarks and gluinos have reached well into the $1-2$ TeV range~\cite{Aaboud:2017vwy,Sirunyan:2018vjp}. Thus, in our discussion we will assume a range of scalar superpartners $M_{SUSY} = M_{3}=\tilde{m}_{q_{1,2,3}} = \tilde{m}_{l_{1,2,3}} = 1 -10$ TeV. For simplicity, we will assume $|\mu| = M_{SUSY}$ in the main results. However, we will comment on other cases in later sections.

\begin{figure}[t]
\centering
	 \includegraphics[scale=1]{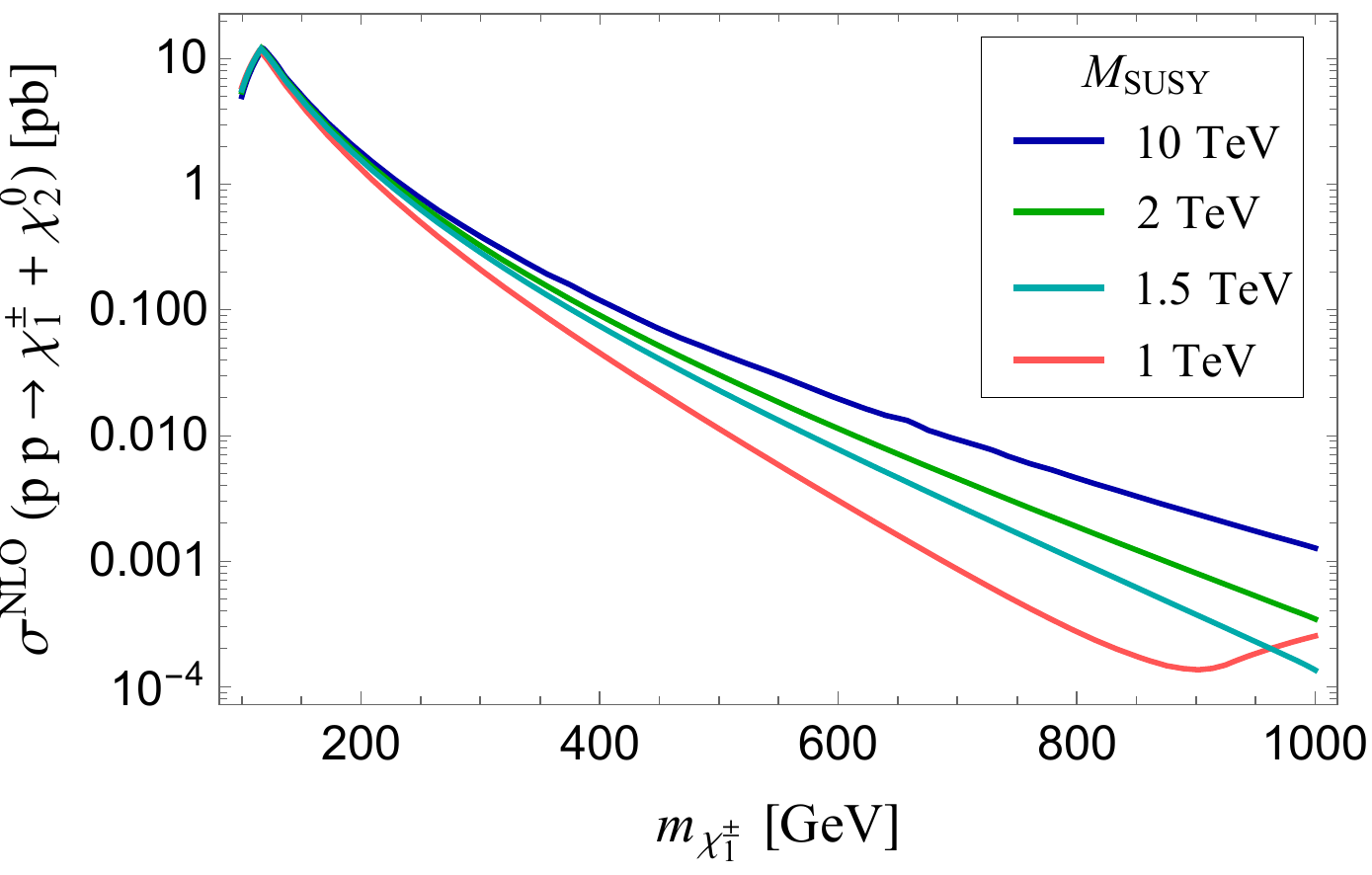} 
\caption{Variation of the Wino-like production cross section for the 13 TeV LHC at NLO when $M_{SUSY}$ is varied between 1 and 10 TeV. $M_{1} = 100$ GeV and $\tan\beta=5$ are fixed and $|\mu|=M_{SUSY}$.}
\label{fig:prod_xsection}
\end{figure}

In Fig. \ref{fig:prod_xsection}, we show the NLO production cross section of Wino-like electroweakinos with respect to the Wino mass for $M_{SUSY}= |\mu| = 1 - 10$ TeV. For large Wino masses, the scalar interactions in the production cross section tend to destructively interfere compared to scenarios when superpartners are decoupled well above the weak scale, with the exception of when $M_{SUSY}=1$ TeV for which the mixing of the Wino with the Higgsinos becomes relevant when $M_{2}$ approaches $\mu$. In the range of $m_{\chi_{1}^{\pm}}\simeq 500 - 1000$ GeV, we find that the difference in the production cross section can be close to a factor of $\sim 2 - 4$. This range of masses is currently in the region of interest of exclusion and/or discovery limits for future searches of electroweakinos at the LHC. Thus, despite being decoupled from the typical searches, the scale of superpartners can have striking consequences on the interpretation of many channels currently being explored.

\begin{figure}[h!]
\centering
	 \includegraphics[scale=0.5]{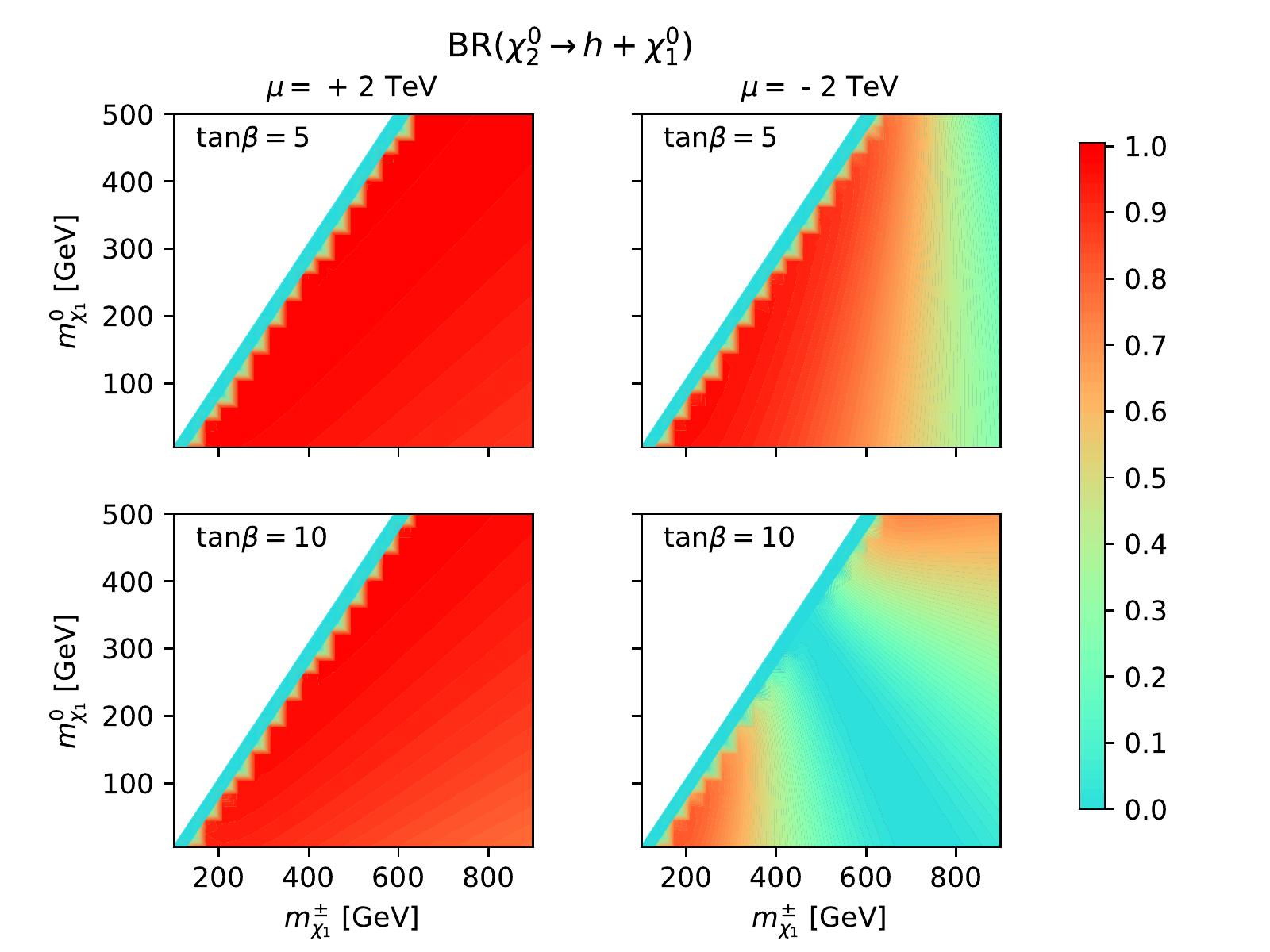}
	  \includegraphics[scale=0.5]{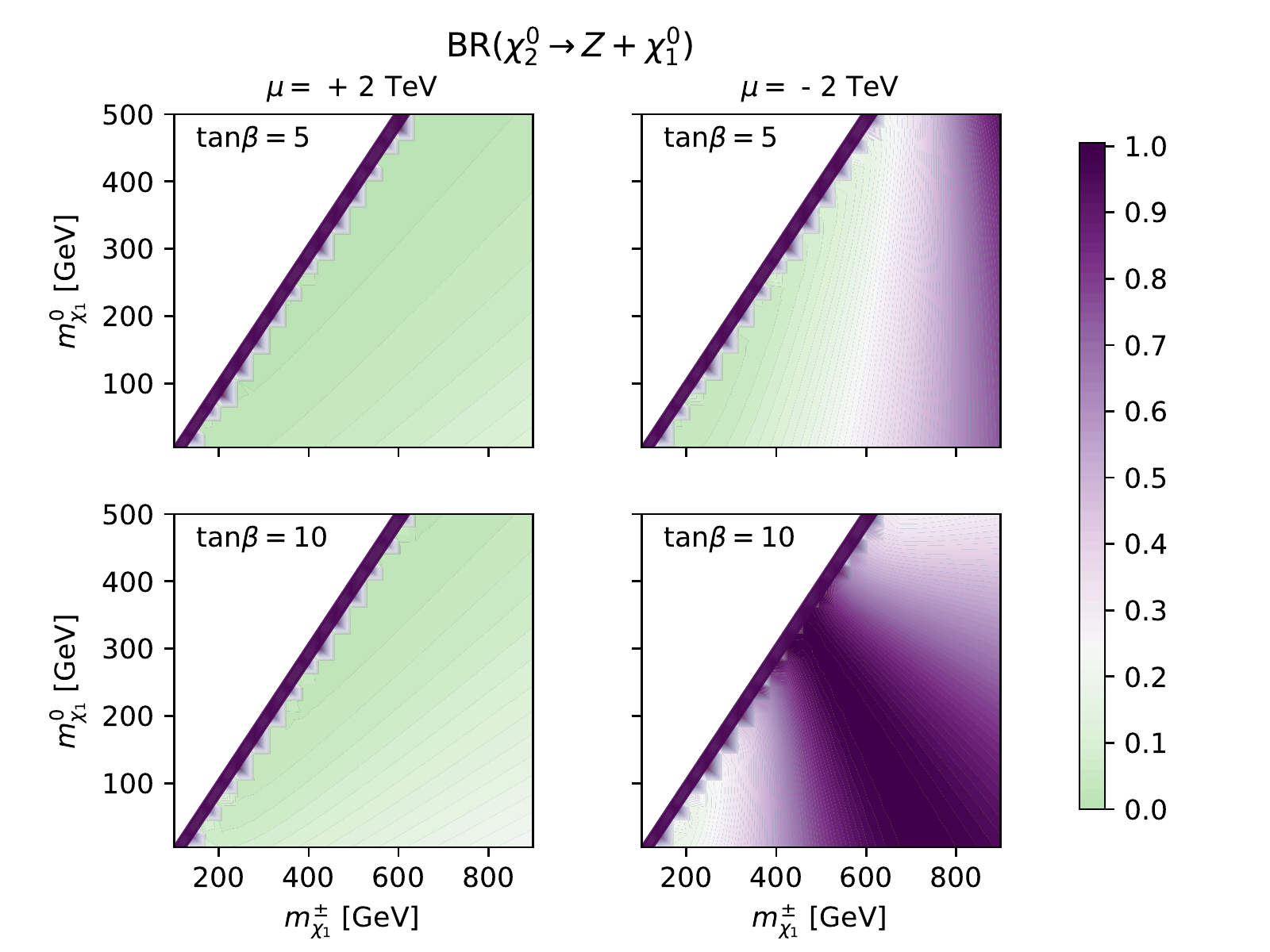} 
\caption{{\bf{Left:}} Branching ratio of $\chi_{2}^{0}$ to the SM Higgs boson and $\chi_{1}^{0}$ for $|\mu| = 2$ TeV and $\tan\beta=5~\&~10$, presented in the $m_{\chi_{1}^{\pm}}$ - $m_{\chi_{1}^{0}}$ plane. {\bf{Right:}} Branching ratio of $\chi_{2}^{0}$ to the Z boson and $\chi_{1}^{0}$ for the same parameters.}
\label{fig:BR_mu_2TeV}
\end{figure}

\begin{figure}[h!]
\centering
	 \includegraphics[scale=0.5]{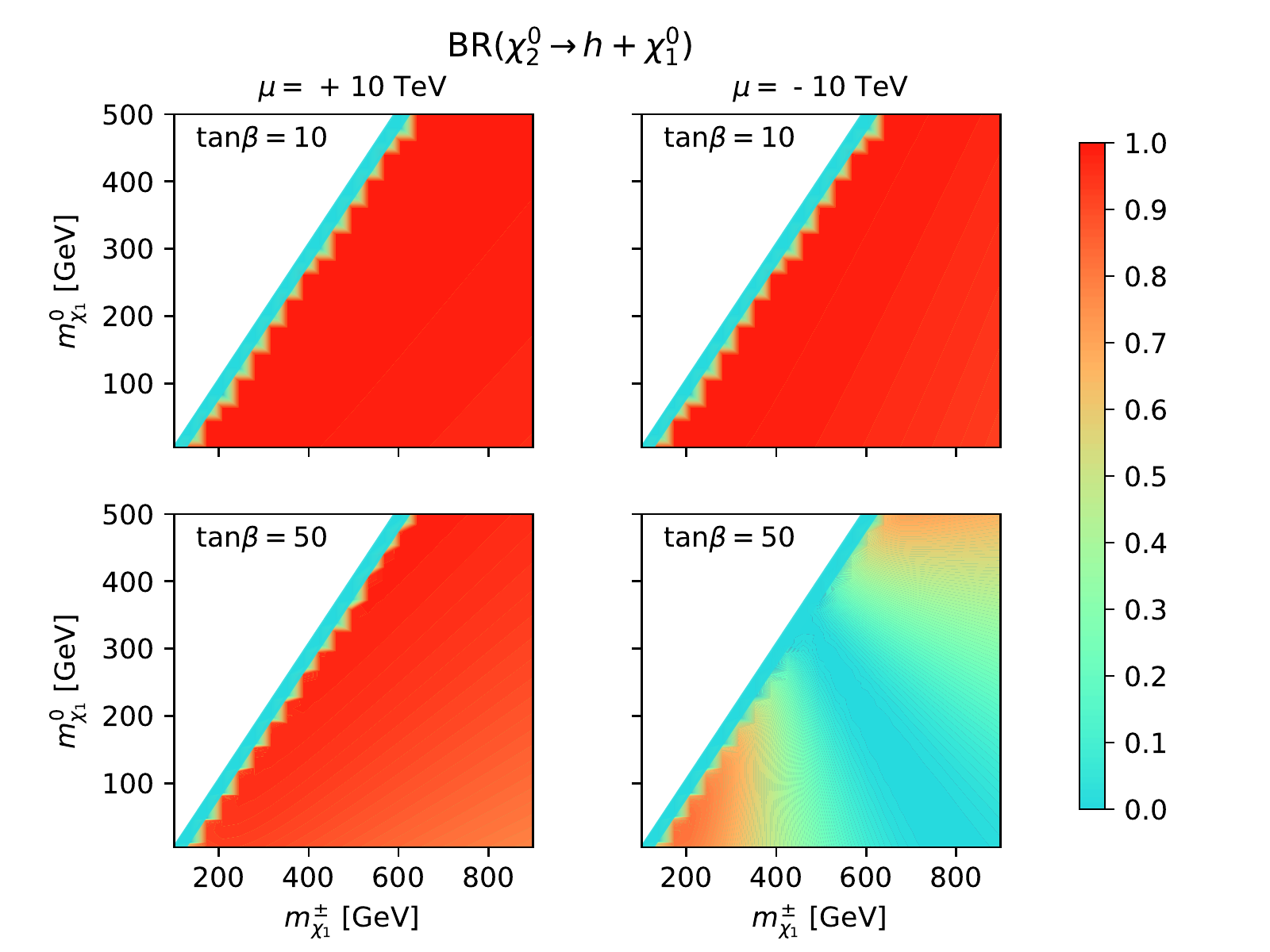}
	  \includegraphics[scale=0.5]{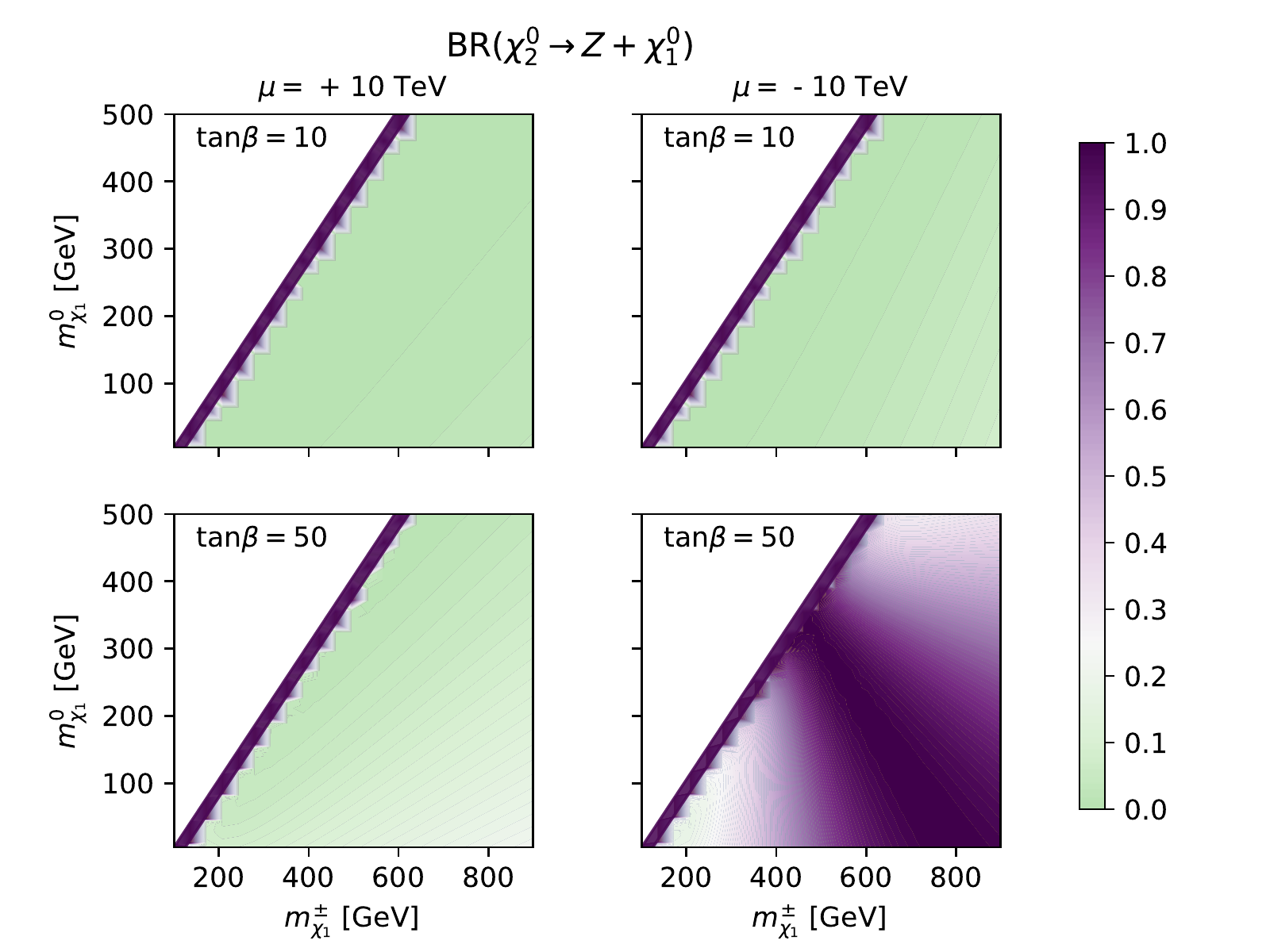} 
\caption{{\bf{Left:}} Branching ratio of $\chi_{2}^{0}$ to the SM Higgs boson and $\chi_{1}^{0}$ for $|\mu| = 10$ TeV and $\tan\beta=10~\&~50$, presented in the $m_{\chi_{1}^{\pm}}$ - $m_{\chi_{1}^{0}}$ plane. {\bf{Right:}} Branching ratio of $\chi_{2}^{0}$ to the Z boson and $\chi_{1}^{0}$ for the same parameters.}
\label{fig:BR_mu_10TeV}
\end{figure}

As discussed in the previous section, the Wino will decay either through a Z or Higgs boson to $\chi_{1}^{0}$. In the traditional searches, these decay modes are considered to be maximal over the whole range of masses considered. However, as we have pointed out these branching ratios have non-trivial dependence on the same set of parameters that determine the masses eigenstates. In Figs. \ref{fig:BR_mu_2TeV} \& \ref{fig:BR_mu_10TeV} we show the branching ratios of $\chi_2^{0}$ into $Z$ and $h$ for $M_{SUSY}=2~\&~10$ TeV respectively. In each case, we show branching ratios for $\mu = \pm M_{SUSY}$. For $M_{SUSY}=2$ TeV we show branching ratios for $\tan\beta = 5~\&~10$, while for $M_{SUSY}=10$ TeV we take $\tan\beta = 10~\&~50$ to show the region of parameters where the blind spot in the Higgs decay is realized. The spectrum and branching ratios are produced using {\tt FeynHiggs} \cite{Bahl:2018qog,Bahl:2017aev,Bahl:2016brp,Hahn:2013ria,Hollik:2014bua,Degrassi:2002fi,Heinemeyer:1998np,Heinemeyer:1998yj} and {\tt SUSY-HIT} \cite{Djouadi:2006bz}, respectively, by scanning $M_{1} = [5,500]$ and $M_{2} = [100,1000]$.

We see that for $M_{SUSY} = 2$ TeV the Higgs decay mode is dominant over most of the region of interest. However, for $\mu < 0$ and $\tan\beta = 10$ we see that the previously discussed blind spot condition may be fulfilled and the Z decay mode becomes dominant. While for $M_{SUSY}=10$ TeV the Higgs decay mode reaches maximum strength in most of the parameter space whereas the decay of Winos through the Z boson is negligible everywhere except in the compressed region, $m_{\chi_{2}^{0}}\simeq m_{Z} + m_{\chi_{1}^{0}}$ and for $\mu < 0$ and $\tan\beta = 50$ in this case. We note that such patterns of the branching ratios are not themselves strictly dependent on the scale of the scalar superpartners. Leaving $\mu = -2$ TeV fixed and increasing the scale of the scalar superpartners will have the dual effect of increasing the production cross section and leaving the blind spot in a range accessible to, for instance, trilepton searches.

These findings show that, except for particular regions of parameter space, the Higgs decay channel stands out as the most promising decay mode for searches of Wino-like electroweakinos at the LHC. In particular, if $\mu > 0$ this decay channel is the only relevant production mode beyond the compressed region, $m_{\chi_{2}^{0}}\simeq m_{Z} + m_{\chi_{1}^{0}}$. This, together with the dependence of the production cross section on masses of the scalar superpartners, gives pertinent information when interpreting current bounds and projecting the future reach of the HL-LHC for electroweakinos.

\section{Reach of electroweakino searches according to SUSY scenarios}
\label{sec:reach}

\subsection{Current bounds}

The current reach of electroweakinos at the LHC has been presented in numerous studies by the ATLAS  and CMS collaborations Ref.~\cite{Aaboud:2018ngk, Aaboud:2018jiw, Aad:2019vvf, ATLAS-CONF-2019-020,Sirunyan:2017zss,Sirunyan:2017lae,Sirunyan:2018ubx,Sirunyan:2018iwl}. For definiteness, in this article we will concentrate on the studies presented by the ATLAS collaboration. The existing searches present bounds on the masses of charginos and neutralinos at the 13 TeV LHC for luminosities ranging from $36\text{ fb}^{-1}$ to $139\text{ fb}^{-1}$ assuming Wino-like crosses sections with superpartners decoupled and maximal branching ratios of $\chi_{2}^{0}$ to either Z or SM Higgs bosons. In this section, we recast the current bounds comparing the reach of electroweakino searches with respect to the scale of superpartners.

\begin{figure}[t]
\centering
	\includegraphics[width=0.43 \columnwidth]{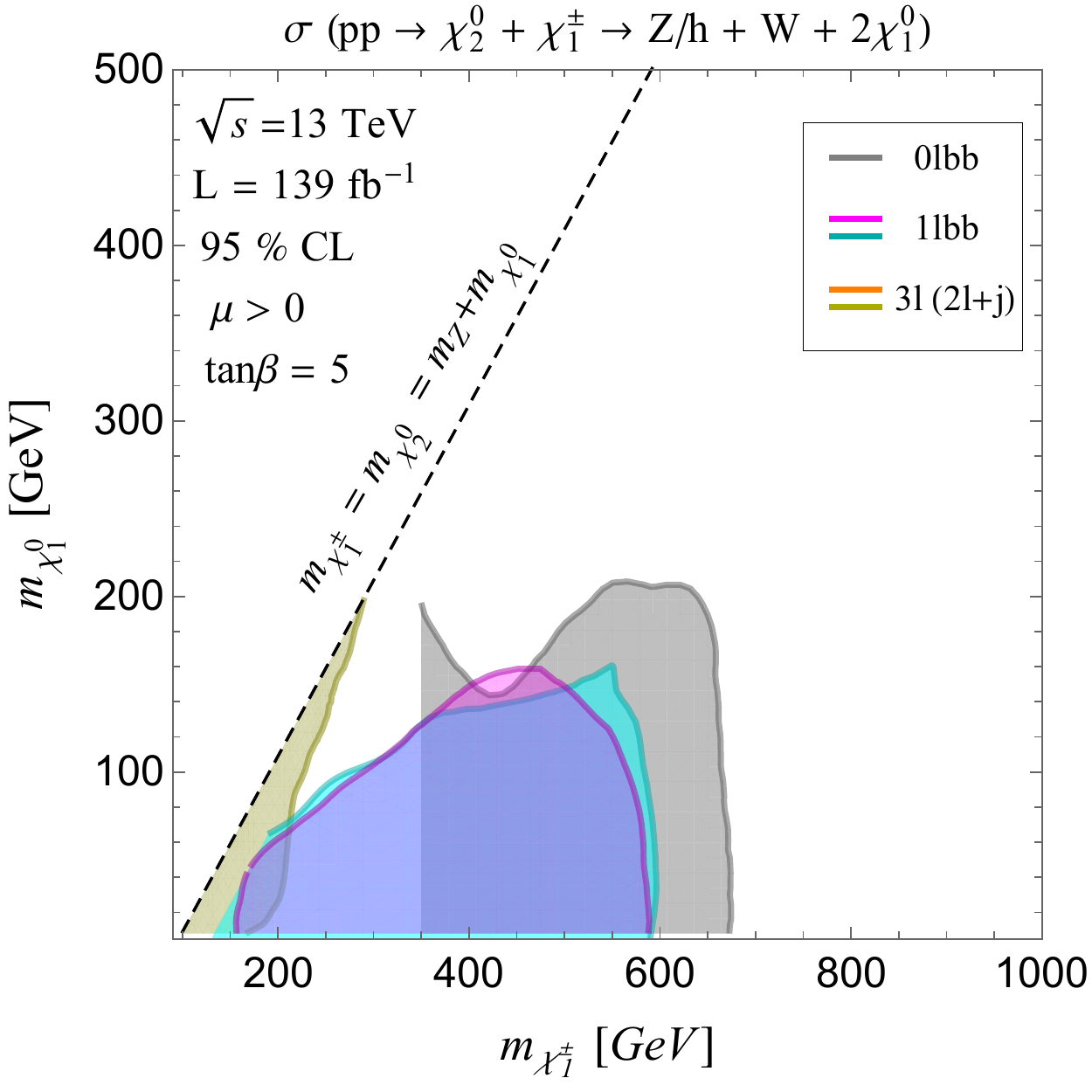} 
	\includegraphics[width=0.43 \columnwidth]{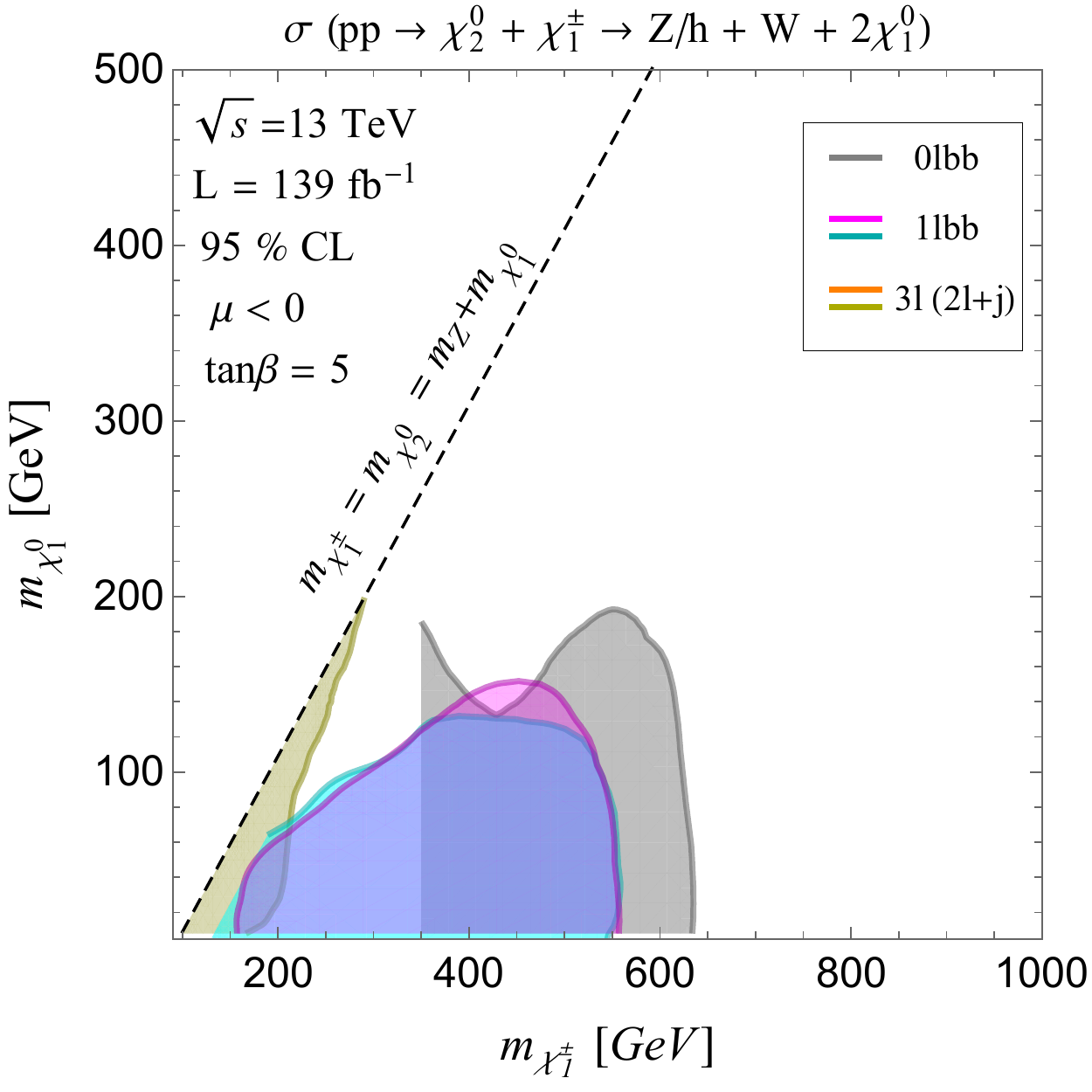} \\
	\includegraphics[width=0.43 \columnwidth]{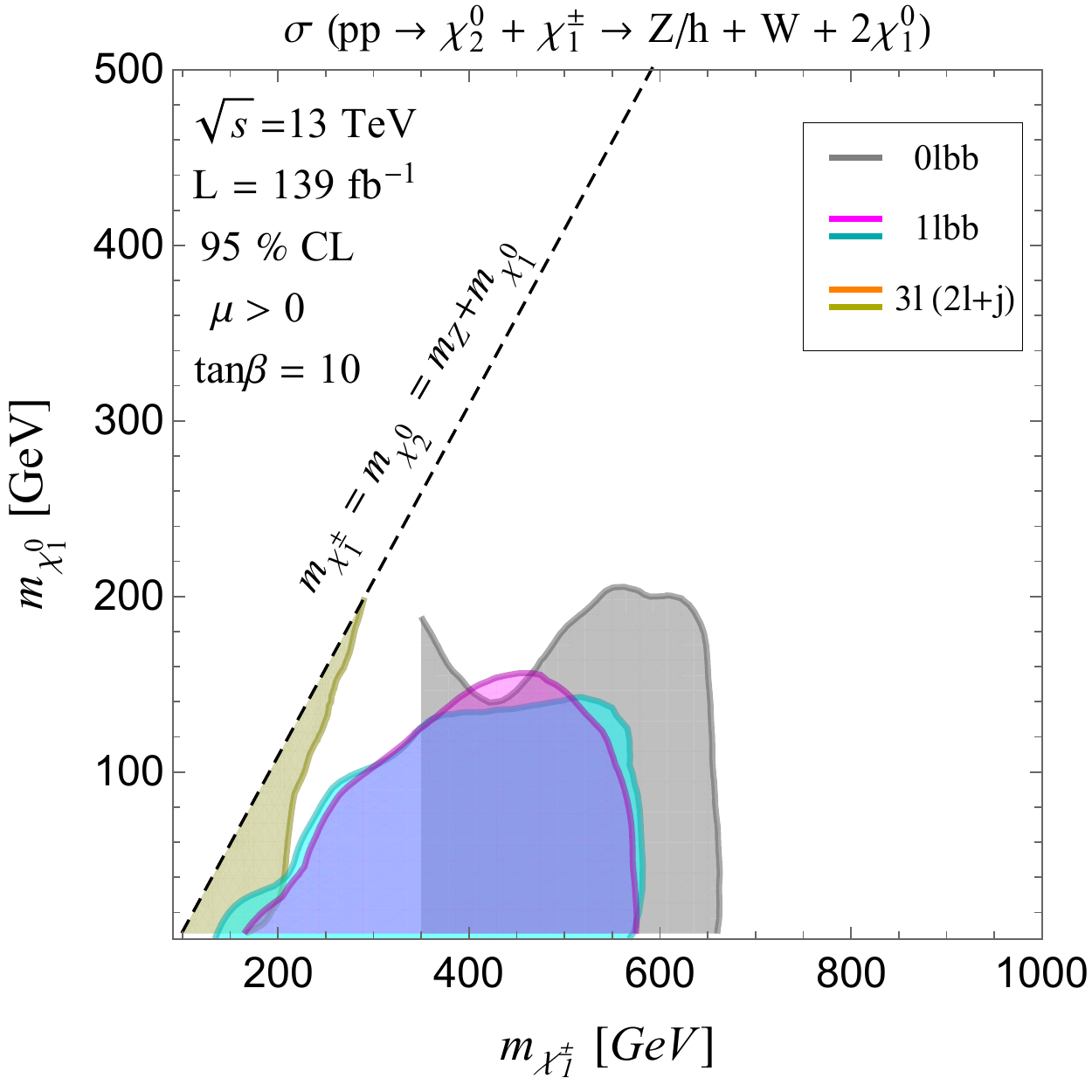} 
	\includegraphics[width=0.43 \columnwidth]{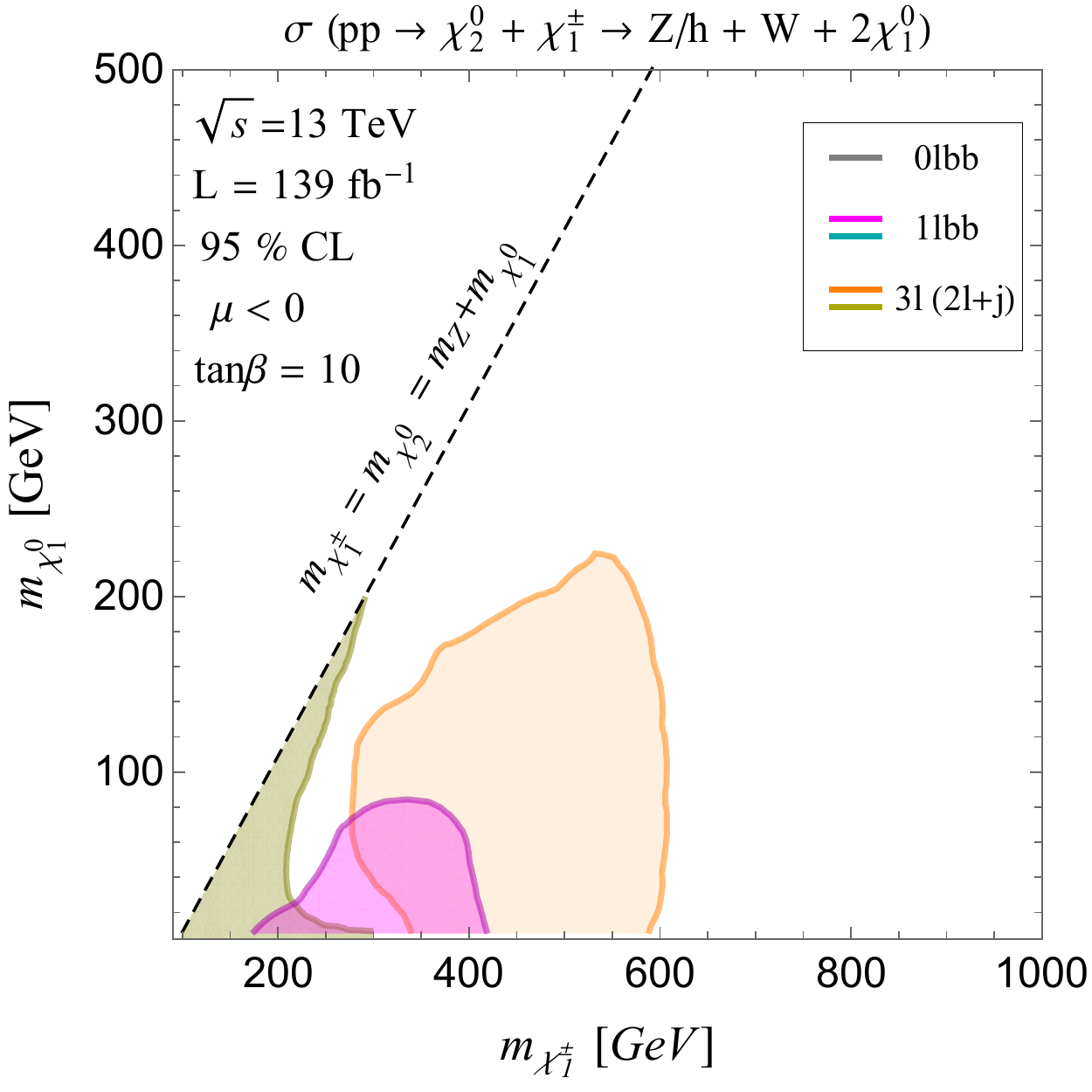} 
\caption{
$95\%$ confidence level bounds on the Wino-Bino scenario projected to integrated luminosity $L=139 ~\text{fb}^{-1}$.
In the top (bottom) panel, we show the bounds for $\tan\beta = 5$ ($\tan\beta = 10$) and $M_{SUSY}=|\mu|=2\text{TeV}$. The $0\ell bb$ (gray)~\cite{Aaboud:2018ngk} and $1\ell bb$ (magenta, cyan)~\cite{Aaboud:2018ngk,Aad:2019vvf} bounds are projected from searches of the $\chi_{2}^0 \chi_1^\pm\to hW+2\chi_1^0$ channel, with $h\to \bar{b}b$ and $W$ decay to hadronic or leptonic final states. The $3\ell$ (dark yellow)~\cite{ATLAS-CONF-2019-020}  and $3\ell/2\ell+\text{j}$ (orange)~\cite{ Aaboud:2018sua}  bounds are projected from the $\chi_{2}^0 \chi_1^\pm\to ZW+2\chi_1^0$ channel, with $Z \to 2\ell$. }
\label{fig:2TeV_constraints}
\end{figure}

The production cross sections and associated branching ratios are obtained over the Wino-Bino parameter space, $M_{1} = [5,500]$ and $M_{2} = [100,1000]$, using and  {\tt Prospino-2.1} \cite{Beenakker:1996ed} and {\tt SUSY-HIT} respectively. For each point in the scan, we compare the result with the experimental upper limits on the cross section for decays leading to trilepton or $h\rightarrow b\bar{b}$ channels~\cite{Aaboud:2018ngk, Aaboud:2018jiw, Aad:2019vvf, ATLAS-CONF-2019-020},

\begin{equation}
pp\rightarrow \chi_{1}^{\pm} + \chi_{2}^{0}\rightarrow W^{\pm} + Z/h + \slashed{E}_{T}= \begin{cases*}
                    3 \ell  + \slashed{E}_{T} & for $Z\rightarrow \ell\ell$\\
                     1(0)\ell + b\bar{b} + \slashed{E}_{T} & for $h\rightarrow b\bar{b}$.
                 \end{cases*}
\end{equation}

\begin{figure}[t]
\centering
	\includegraphics[width=0.43 \columnwidth]{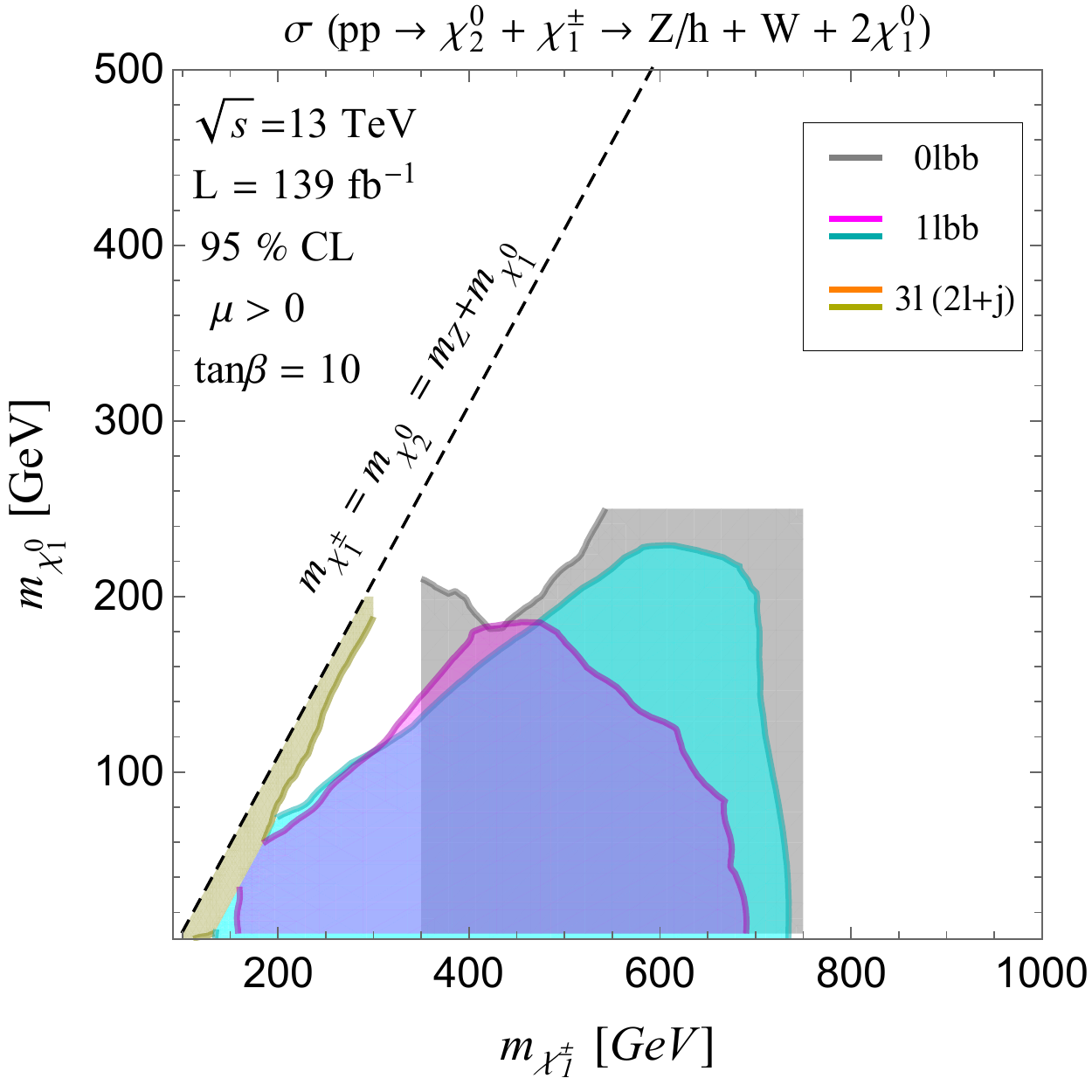} 
	\includegraphics[width=0.43 \columnwidth]{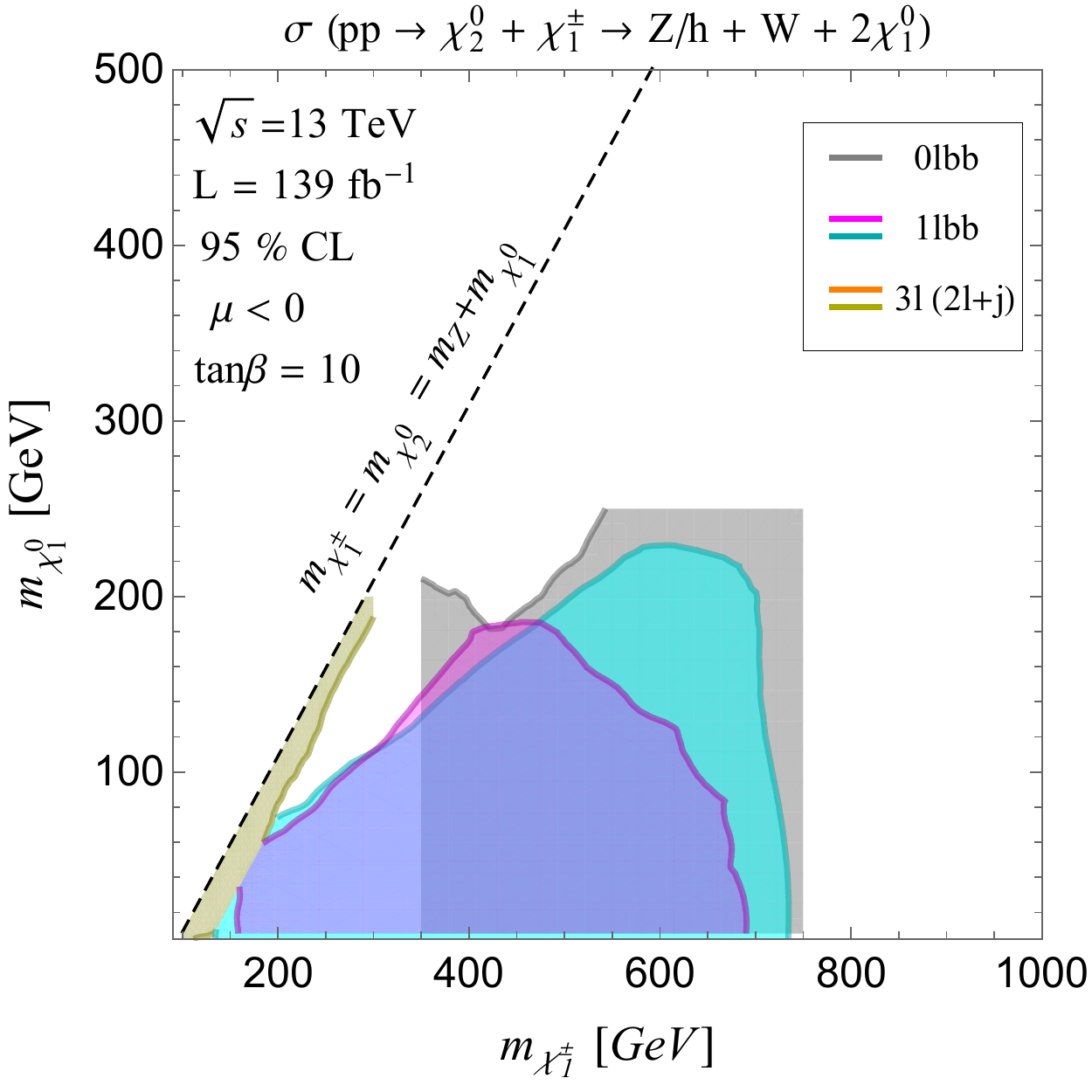}\\ 
	\includegraphics[width=0.43 \columnwidth]{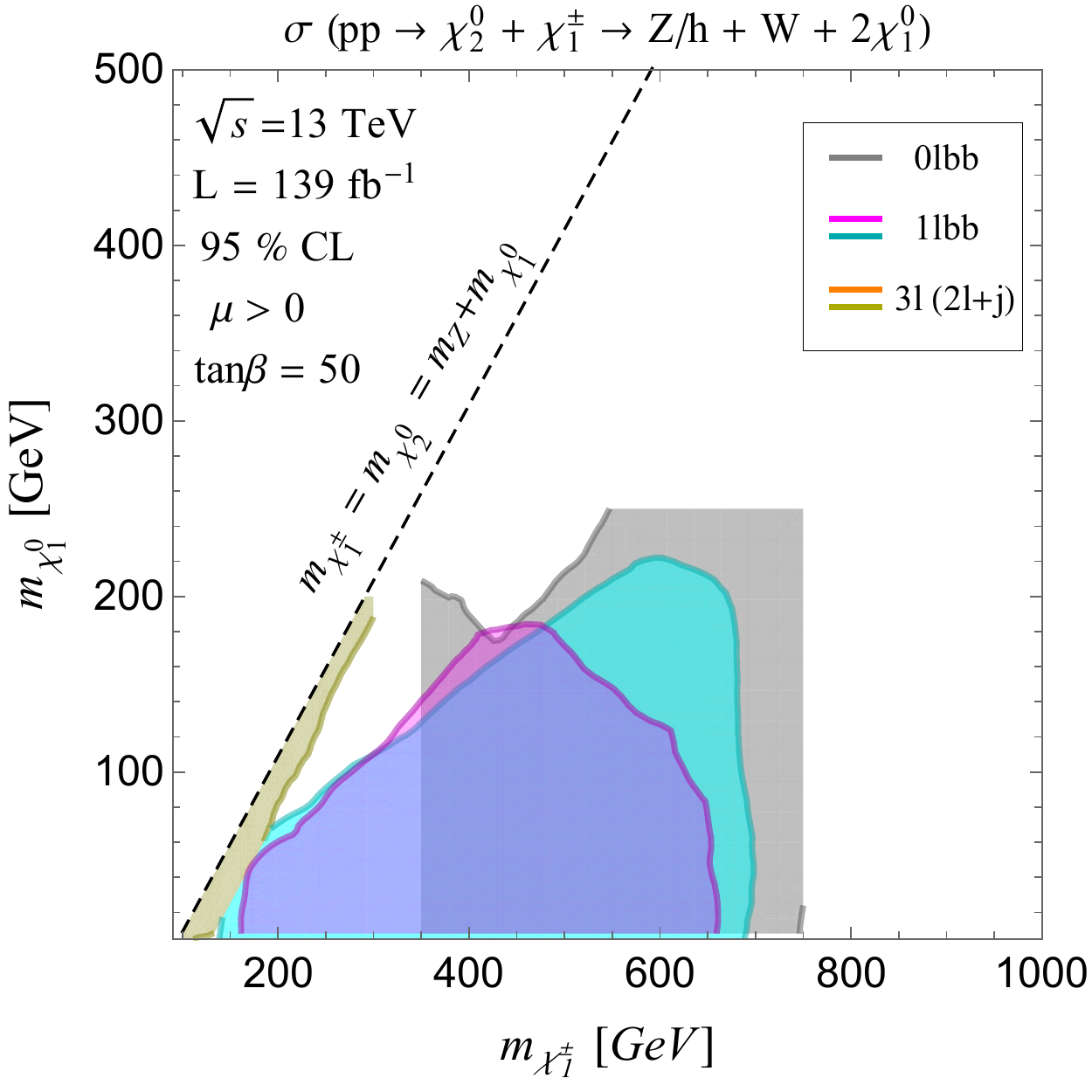} 
	\includegraphics[width=0.43 \columnwidth]{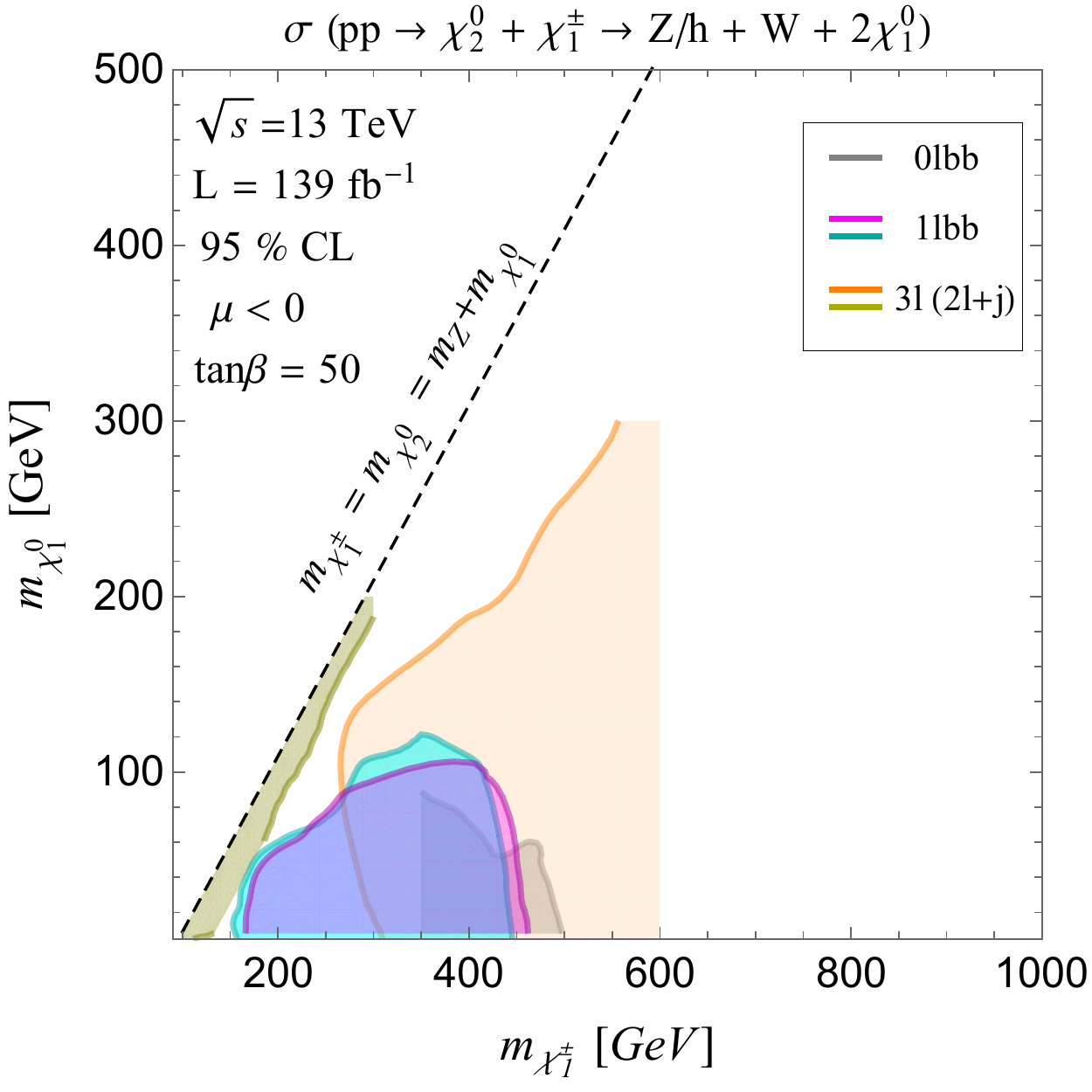}\\ 
\caption{ $95\%$ confidence level bounds on the Wino-Bino scenario projected for integrated luminosity of $L=139~\text{fb}^{-1}$ with $M_{SUSY}=|\mu|=10$ TeV. The labels are similar to Fig.~\ref{fig:2TeV_constraints}.}
\label{fig:10TeV_constraints}
\end{figure}

In this and subsequent sections, we focus mainly on two scenarios when $M_{SUSY}=|\mu| = 2~\text{TeV}$ and $M_{SUSY}=|\mu| = 10$ TeV.~\footnote{Lower squark masses $~1$ TeV would result in abysmal reach for electroweakinos as the interference in the production cross section is maximal.} As discussed in the previous section, for a given $|\mu|$ the blind spot in the Higgs decay occurs for different regions of $\tan\beta$. Thus, in order to present both the worst and best case scenario for the reach of the searches we consider $\tan\beta = 5~\&~10$ when $M_{SUSY}=|\mu| = 2$ TeV, and $\tan\beta = 10~\&~50$ when $M_{SUSY}=|\mu| = 10$ TeV. In each case, since we are recasting the current bounds taken from the data available on {\tt HEP-data}, we do not extend our results beyond what has already been explored by the experimental collaborations as such an analysis would require an artificial presentation of signal and background efficiencies which would typically improve with increased data. Thus, we consider our results conservative in this sense.

\begin{figure}[t]
\centering
	\includegraphics[width=0.45 \columnwidth]{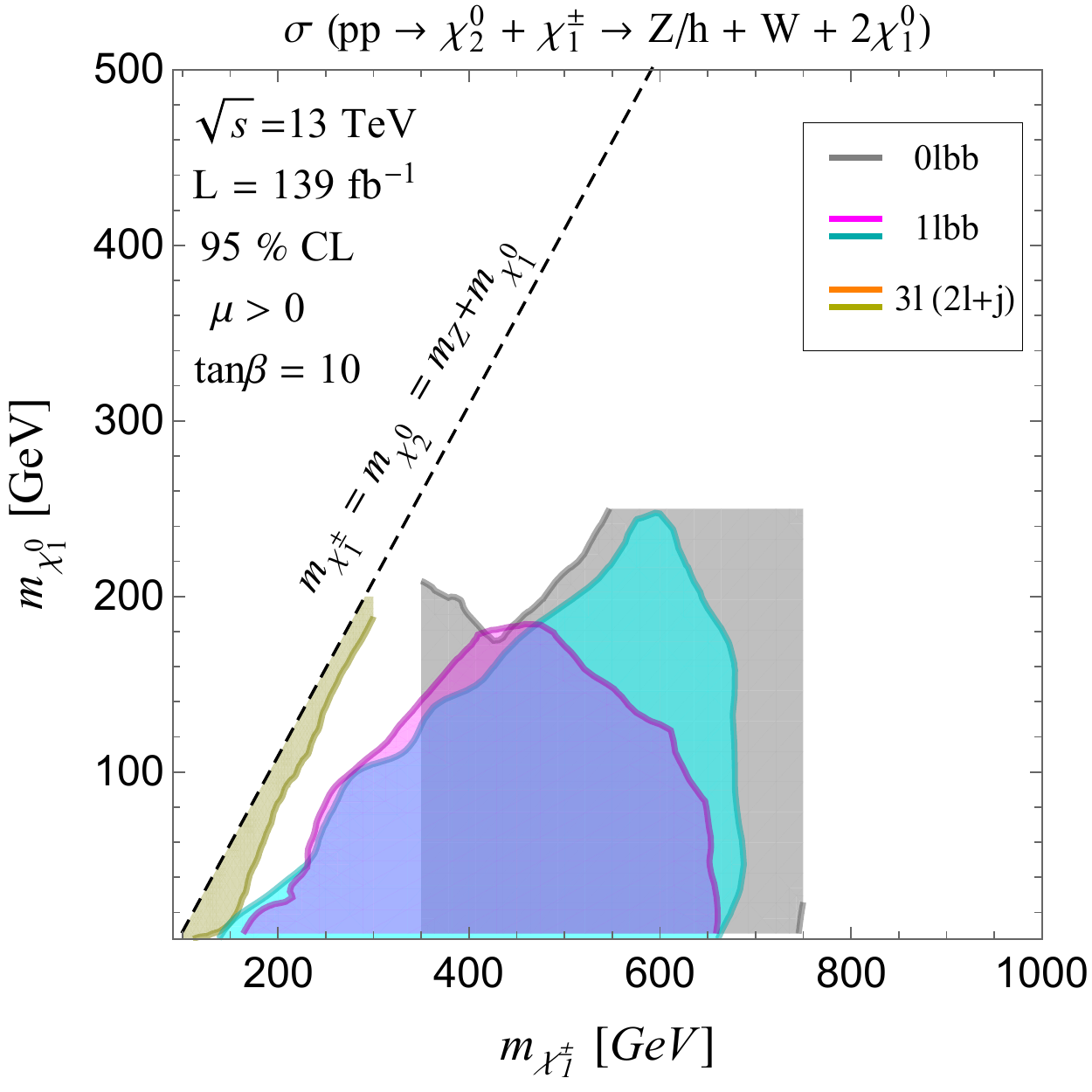} 
	\includegraphics[width=0.45 \columnwidth]{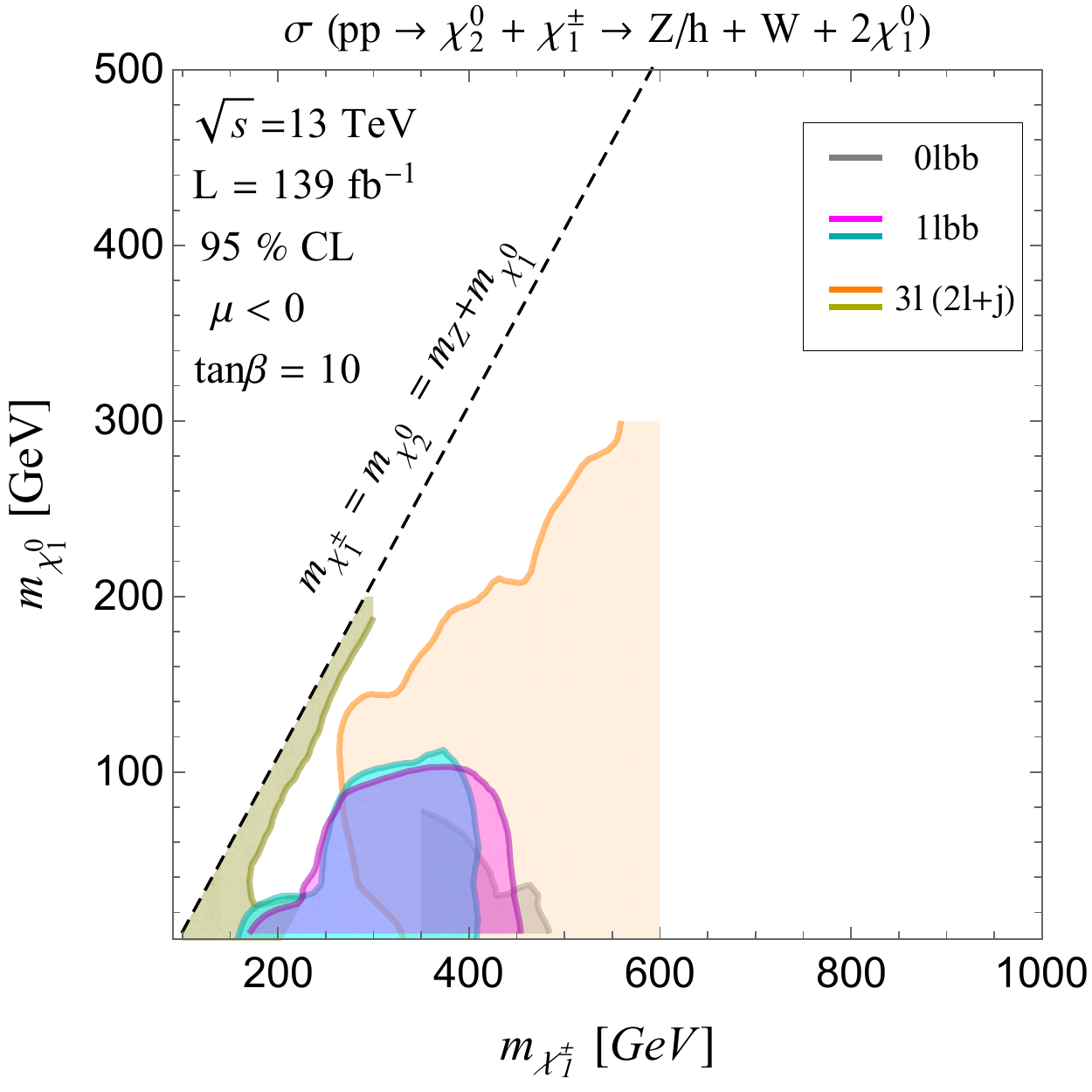} 
\caption{ $95\%$ confidence level bounds on the Wino-Bino scenario projected to integrated luminosity $L=139 ~\text{fb}^{-1}$, assuming $\tan\beta = 10$, $|\mu|=2$ TeV, and $M_{SUSY}=10$ TeV. The labels are similar to Fig.~\ref{fig:2TeV_constraints}.}
\label{fig:split_constraints}
\end{figure}

In Fig.~\ref{fig:2TeV_constraints}, we show the result of recasting the current bounds for $M_{SUSY}=|\mu| = 2$ TeV. The gray and magenta shaded regions show the bounds from a search of the decay $\chi_{2}^0 \chi_1^\pm\to hW+2\chi_1^0$, followed by $h\rightarrow b\bar{b}$, in final states  with a pair $b$-quarks and zero or one leptons respectively \cite{Aaboud:2018ngk}. We have denoted these channels as $0\ell bb$ and $1\ell bb$. The cyan region shows the corresponding bound from an independent search of the $1\ell bb$ channel \cite{Aad:2019vvf}. The orange shaded region, denoted as $3\ell(2\ell+j)$, shows our recasting of the bounds resulting from a search of the $\chi_{2}^0 \chi_1^\pm\to ZW+2\chi_1^0$ decay, followed by $Z\rightarrow \ell^{+}\ell^{-}$, in the statistical combination of final states with 3 leptons and 2 leptons plus jets \cite{Aaboud:2018jiw}. Finally, the dark yellow region shows the bounds from a similar trilepton search focused in the compressed electroweakino spectrum~\cite{ATLAS-CONF-2019-020}.

Due to the decrease in the production cross section, the resulting bounds from the Higgs channel are  weaker than what is typically presented, reaching slightly above chargino masses $m_{\chi_{1}^{\pm}} \simeq 600$ GeV and neutralino masses $m_{\chi_{1}^{\pm}} \simeq 200$ GeV. Meanwhile the trilepton searches lose sensitivity almost over the whole range of masses, except in the region of parameters where the blind spot appears in the Higgs decay mode, see Fig. \ref{fig:BR_mu_2TeV}. In contrast, when $M_{SUSY}=|\mu| = 10$ TeV, Fig.~\ref{fig:10TeV_constraints}, we find similar reach in the Higgs channel as is currently expected (in this and subsequent figures, regions of parameters that were left unexplored by the experimental analyses are shown at the edge of the bounds by sharp edges without solid lines). The trilpeton searches again lose sensitivity everywhere beyond the compressed region, and except when $\mu <0$ and $\tan\beta = 50$ due to the suppression of the branching ratio of $\chi_{2}^{0}$ to $Z$. However, in this case the overall reach also improves due to the increase in the production cross section.

The assumption that $M_{SUSY}=|\mu|$ places strong constraints on both the production cross section, patterns of decays for the lightest electroweakino states, and thus the resulting bounds. Other scenarios with different hierarchies, such as in Split Supersymmetry~\cite{Wells:2003tf,ArkaniHamed:2004fb,Giudice:2004tc} where sfermions are much heavier than the gauginos, are also well motivated. In Fig.~\ref{fig:split_constraints} we show the bounds for $\tan\beta = 10$, $|\mu|=2$ TeV, and $M_{SUSY}=10$ TeV. The overall effect on the bounds is twofold. The decoupling of scalars gives an increase in the production cross section yeilding a slightly larger reach in both the Higgs and trilepton channels, and for negative $\mu$ the trilepton searches remain sensitive to the blind spot resulting in an even larger reach compared to the case of a universal SUSY scale (bottom right panel of  Fig.~\ref{fig:2TeV_constraints}).

\begin{figure}[h!]
\centering
	\includegraphics[width=0.43 \columnwidth]{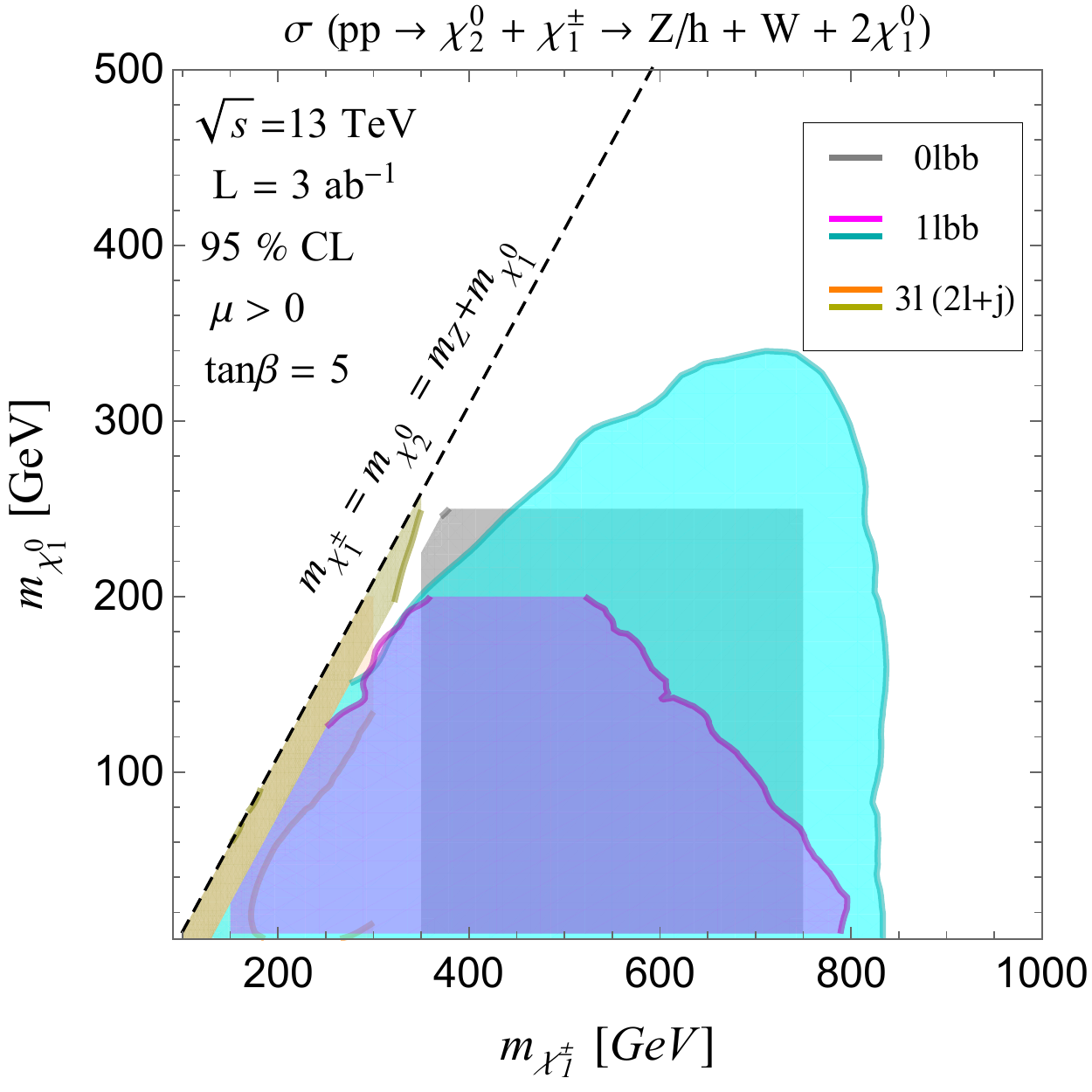} 
	\includegraphics[width=0.43 \columnwidth]{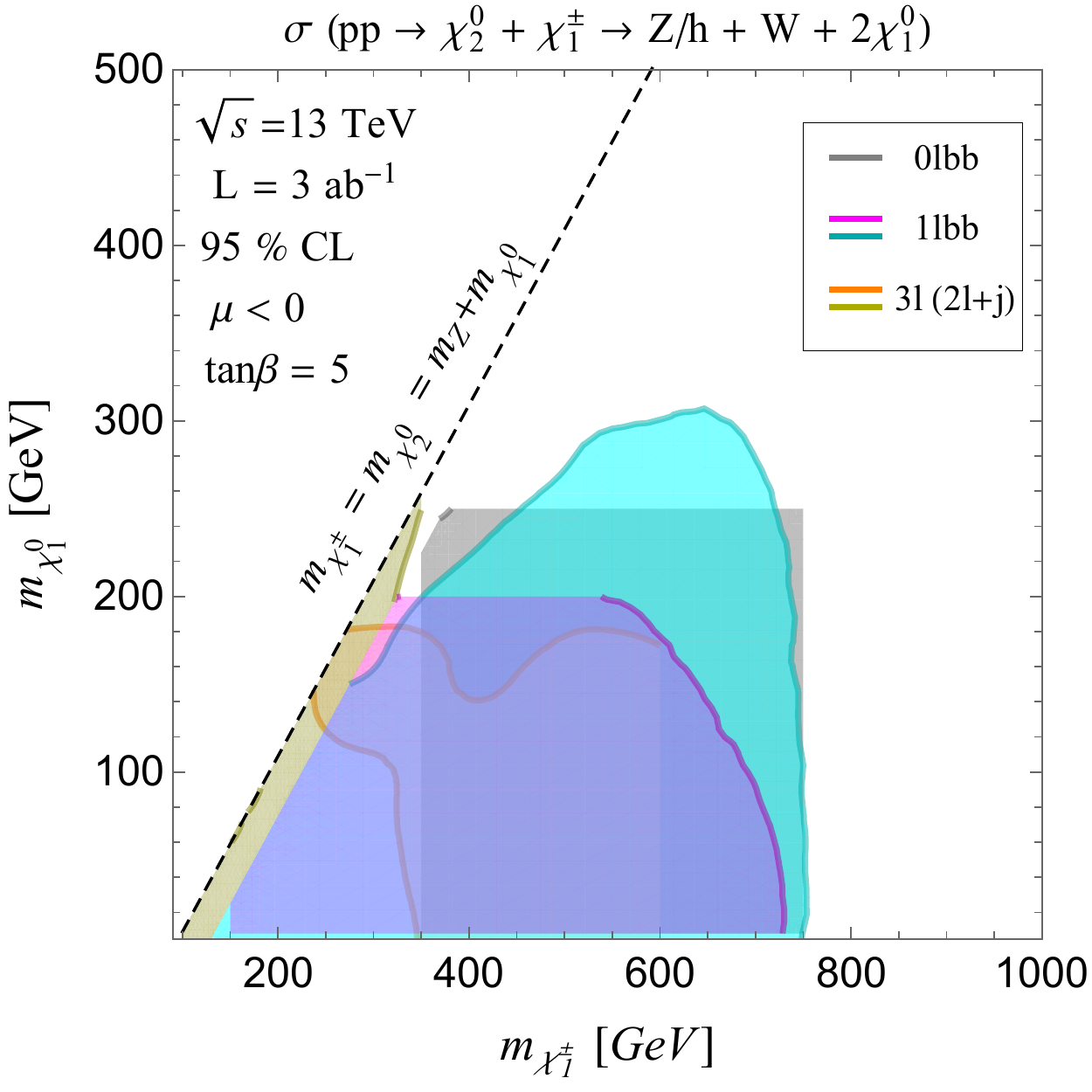} \\
	\includegraphics[width=0.43 \columnwidth]{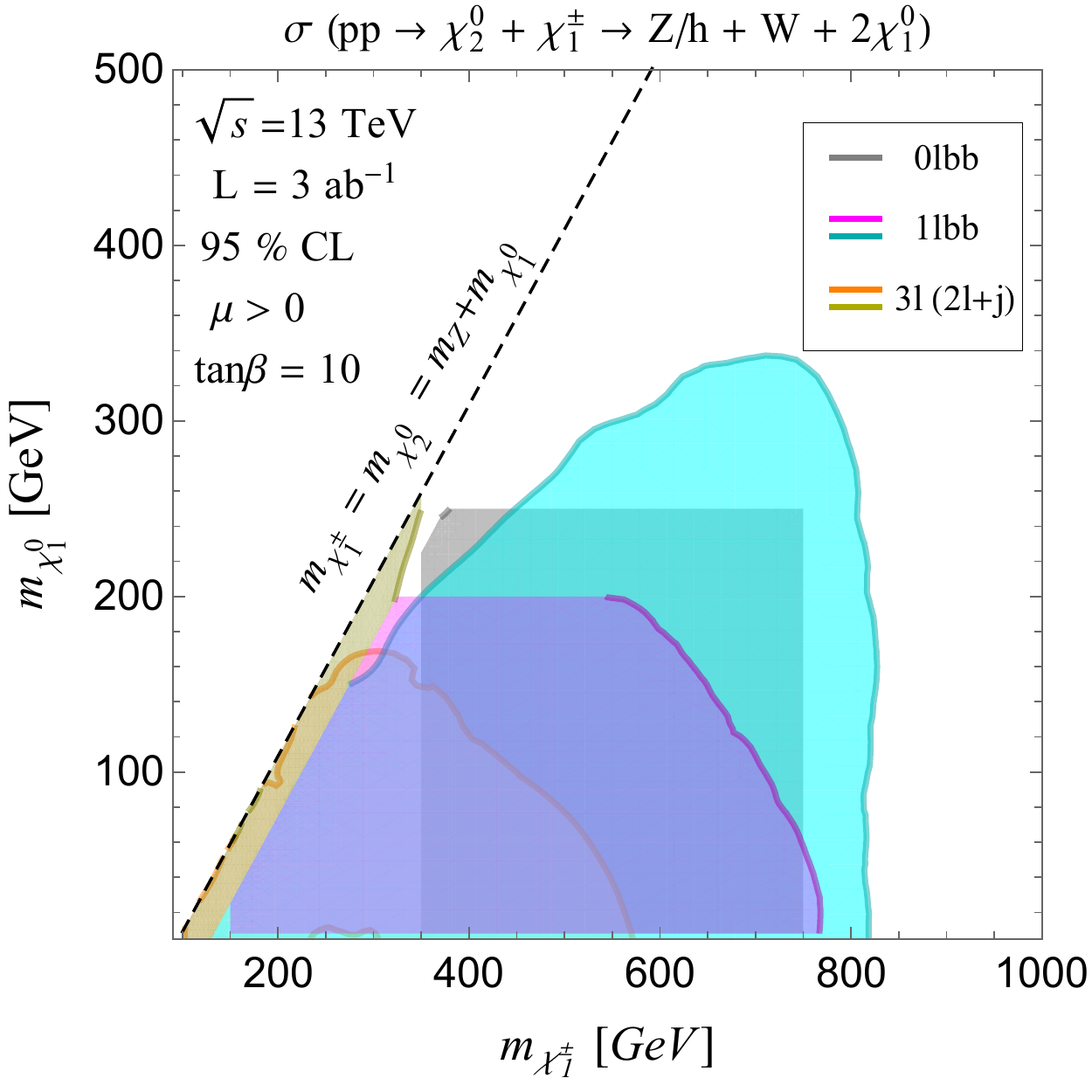} 
	\includegraphics[width=0.43 \columnwidth]{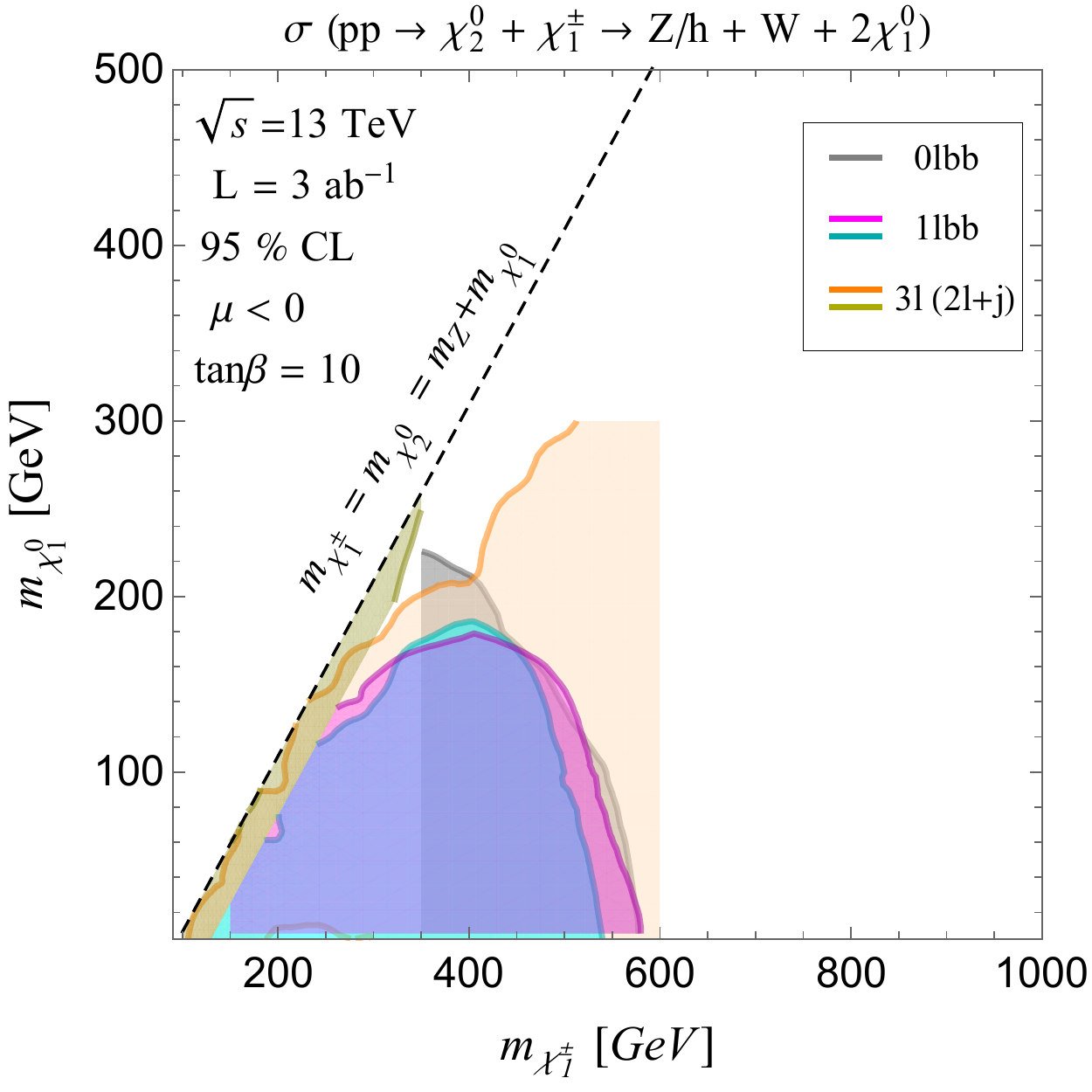} 
\caption{ $95\%$ confidence level bounds on the Wino-Bino scenario projected for integrated luminosity of $L =3 \text{ab}^{-1}$ with $M_{SUSY}=|\mu|=2$ TeV. The labels are similar to Fig.~\ref{fig:2TeV_constraints}.}
\label{fig:2TeV_constraints-3ab}
\end{figure} 

\subsection{Future reach and discovery potential}

In this section, we assess the ultimate reach and discovery potential of Wino-like electroweakinos at the HL-LHC. As in the previous section, we show the projected bounds resulting from the dependence of the scale of superpartners, $|\mu|$, and $\tan\beta$. Throughout this section, we stress the ultimate reach of electroweakino searches with respect to the squark masses. The current bounds shown in the previous section suggest electroweakino masses well above 100 GeV. From Fig.~\ref{fig:prod_xsection} it is clear that in this region of parameters, squarks with masses of 10 TeV can boost the Wino cross section by close to a factor of 2 compared to 2 TeV. Thus, in this section we project the bounds and discovery potential of electroweakinos for the HL-LHC for $M_{SUSY}=2$ and 10 TeV. The lower bound, $M_{SUSY}=2$ TeV, is chosen to conservatively satisfy current bounds on squark masses, while $M_{SUSY}=10$ TeV gives the maximum boost to the electroweakino production cross section leading to the strongest potential reach for these particles. For other theoretical projections of electroweakino searches in the (N)MSSM see  \cite{Yu:2014mda,Han:2016qtc, Dutta:2014hma, Han:2014nba, Baum:2019uzg, Barman:2020vzm}.

\begin{figure}[h!]
\centering
	\includegraphics[width=0.43 \columnwidth]{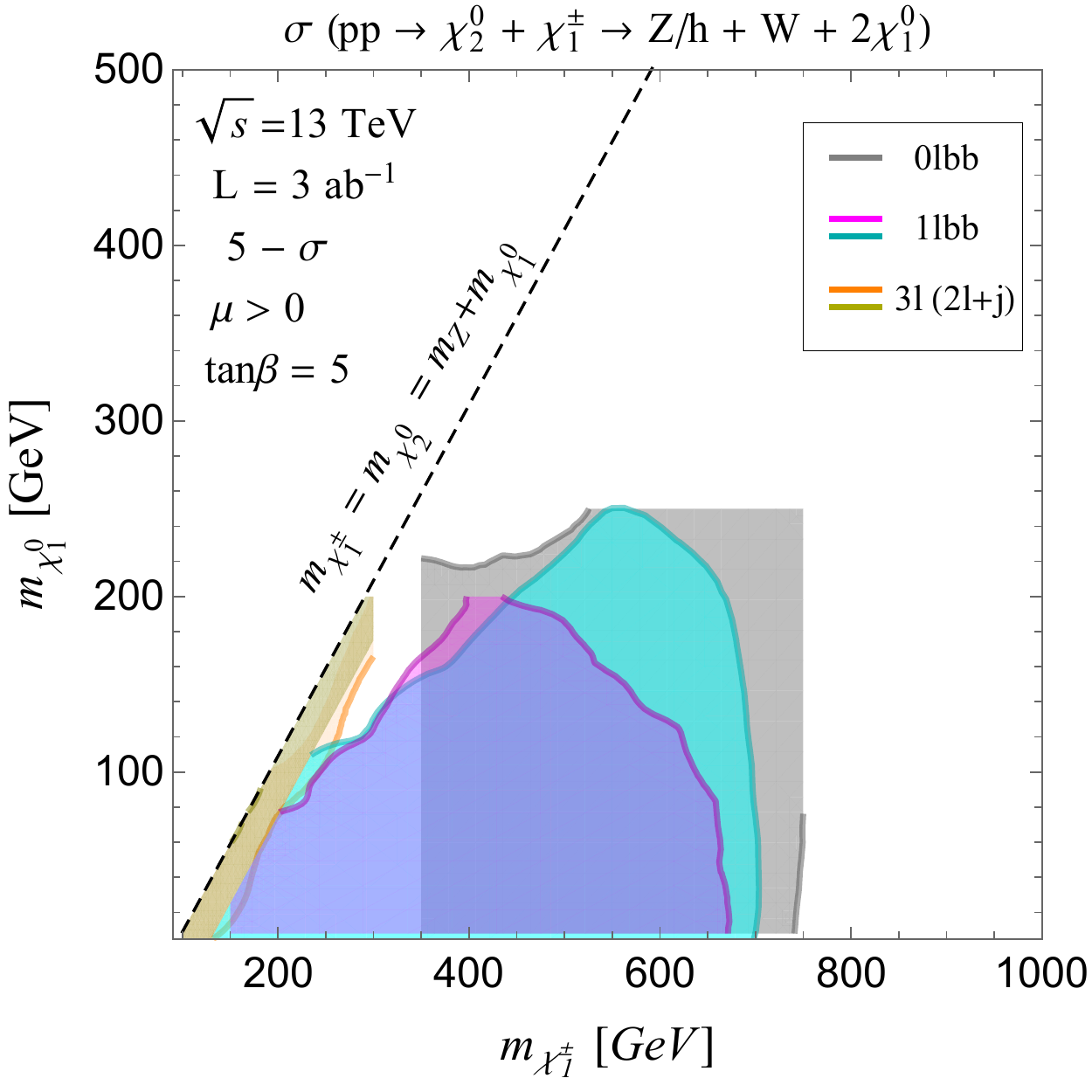} 
	\includegraphics[width=0.43 \columnwidth]{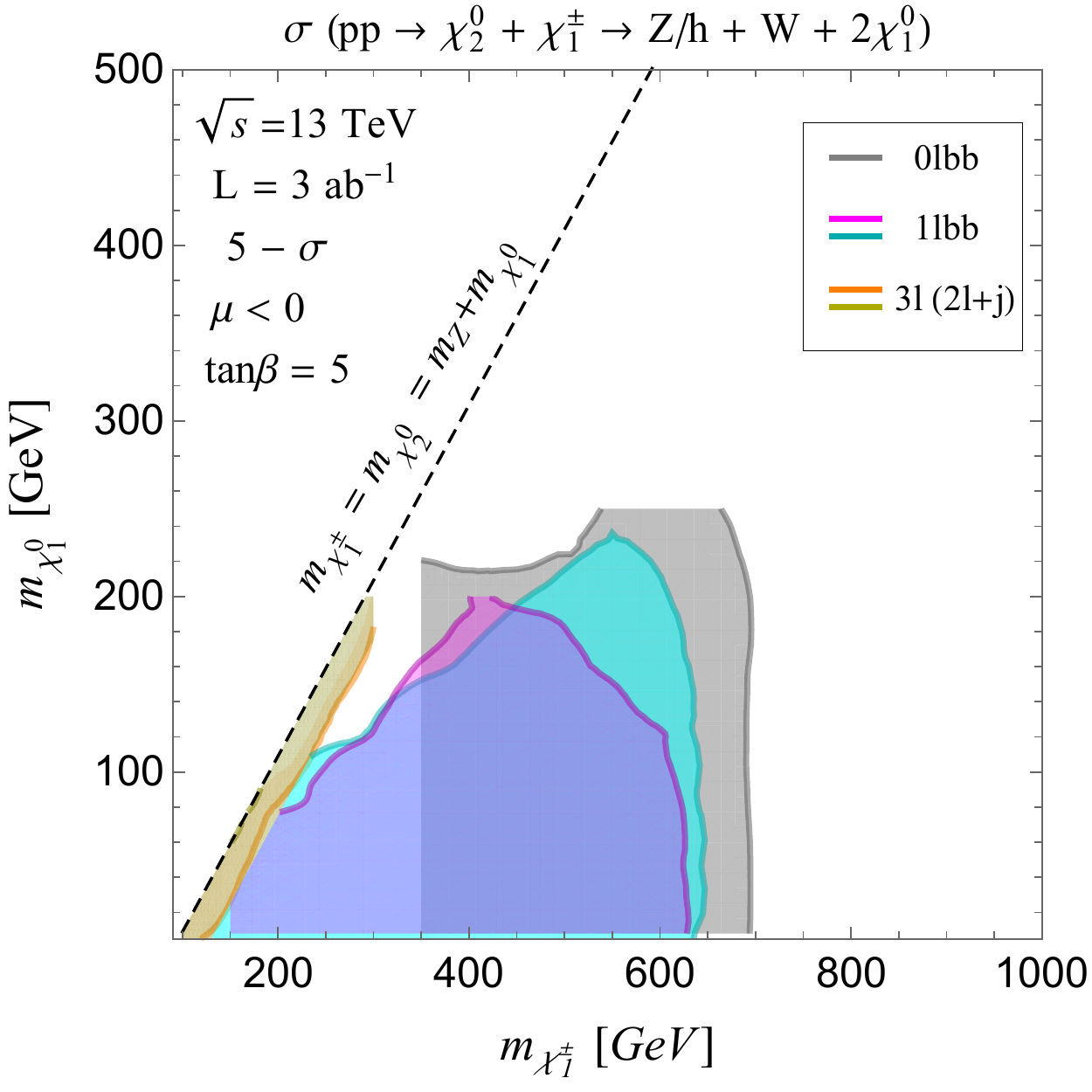} \\
	\includegraphics[width=0.43 \columnwidth]{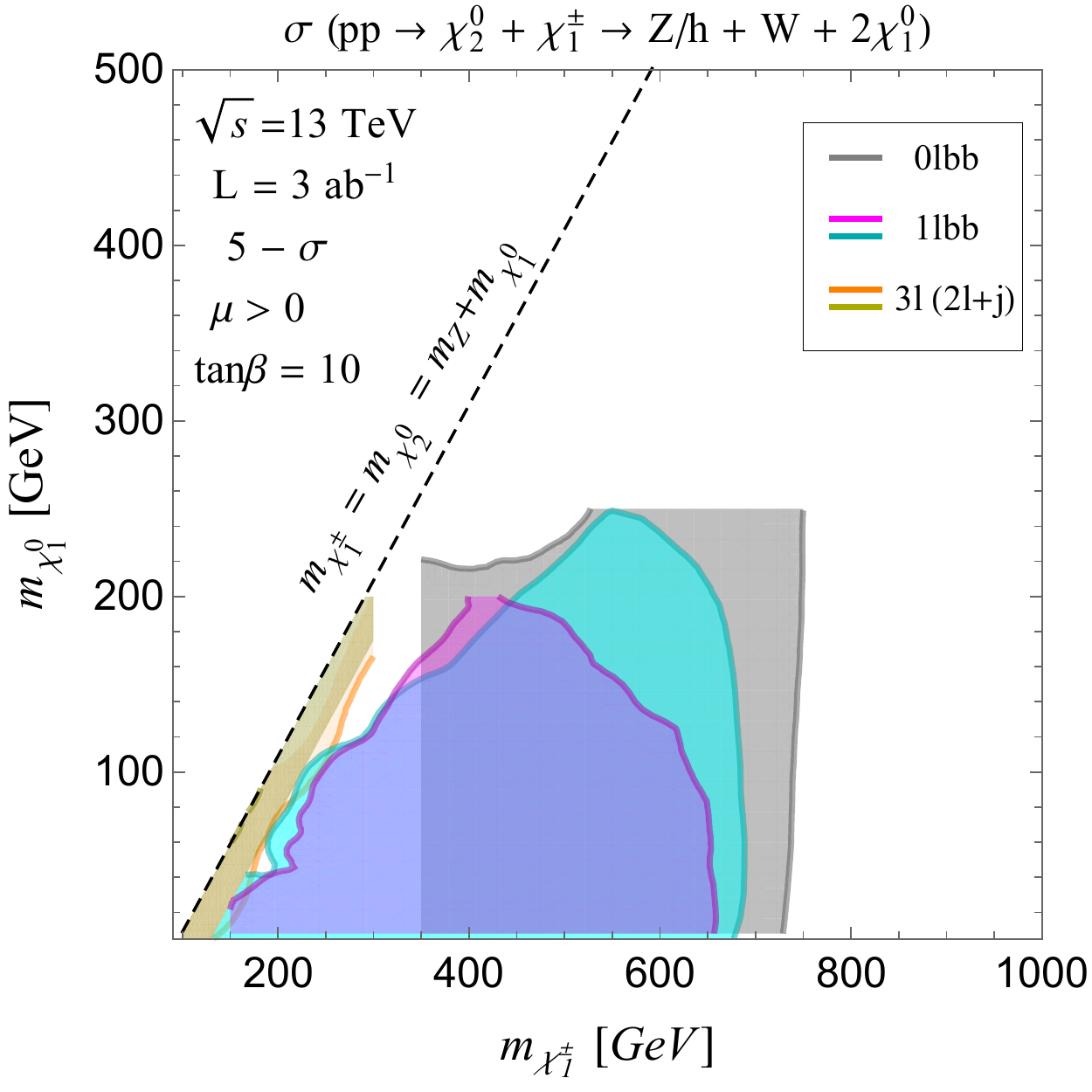} 
	\includegraphics[width=0.43 \columnwidth]{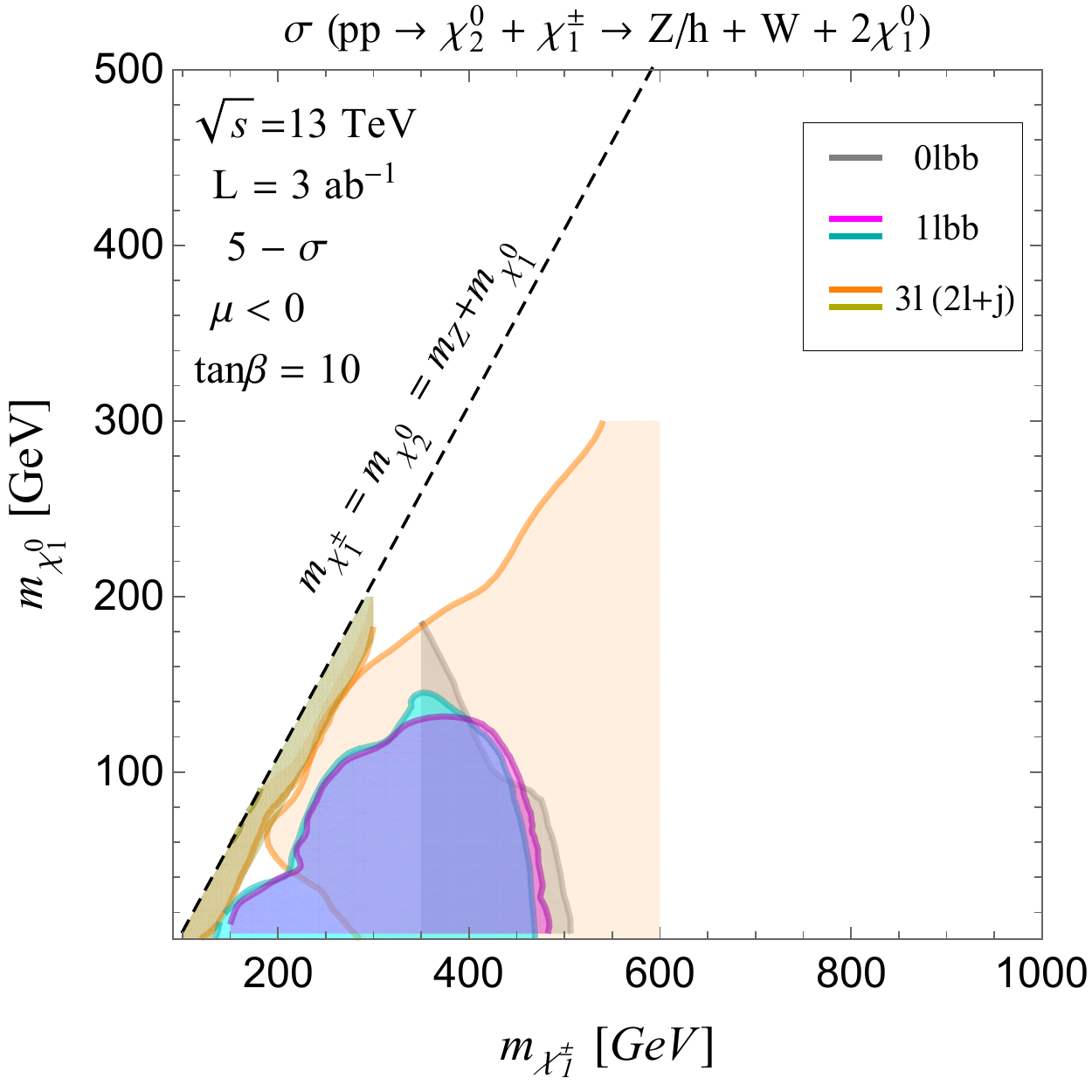} 
\caption{The future potential on the Bino-Wino scenario projected for $5~\sigma$ reach at 3 ab$^{-1}$ with $M_{SUSY}=|\mu|=2$ TeV. The labels are similar to Fig.~\ref{fig:2TeV_constraints}.}
\label{fig:2TeV_constraints-5sigma}
\end{figure}

In Fig.~\ref{fig:2TeV_constraints-3ab}, we show the projected bounds for integrated luminosity of $3~\text{ab}^{-1}$ with $M_{SUSY}=|\mu|=2$ TeV and $\tan\beta = 5~(10)$ in the top (bottom) panels. As before, the Higgs decay channel remains the dominant search channel for most of the range of chargino and neutralino masses, with the ultimate reach extending the bound for $m_{\chi_1^\pm}$ beyond 850 GeV and $m_{\chi_1^{0}}$ to almost 400 GeV when $\mu > 0 $ and $\tan\beta = 5$. However, we see again that these conclusions differ significantly when the coupling of $\chi_{2}^{0}$ to the SM Higgs crosses the blind spot. In this case, the trilepton channel dominates covering a similar range of masses. In Fig.~\ref{fig:2TeV_constraints-5sigma}, we show the $5-\sigma$ discovery regions for the same set of parameters. Comparing each panel to the respective bound in Fig. \ref{fig:2TeV_constraints} we see that there is a significant region of masses in the discovery region at the HL-LHC not excluded by the current searches. Such a region includes $m_{\chi_1^\pm} \gtrsim 600$ GeV, $m_{\chi_1^0} \gtrsim 200$ GeV, and a far better coverage of the compressed region $m_{\chi_1^\pm} - m_{\chi_1^0} \simeq m_Z$.

\begin{figure}[H]
\centering
	\includegraphics[width=0.43 \columnwidth]{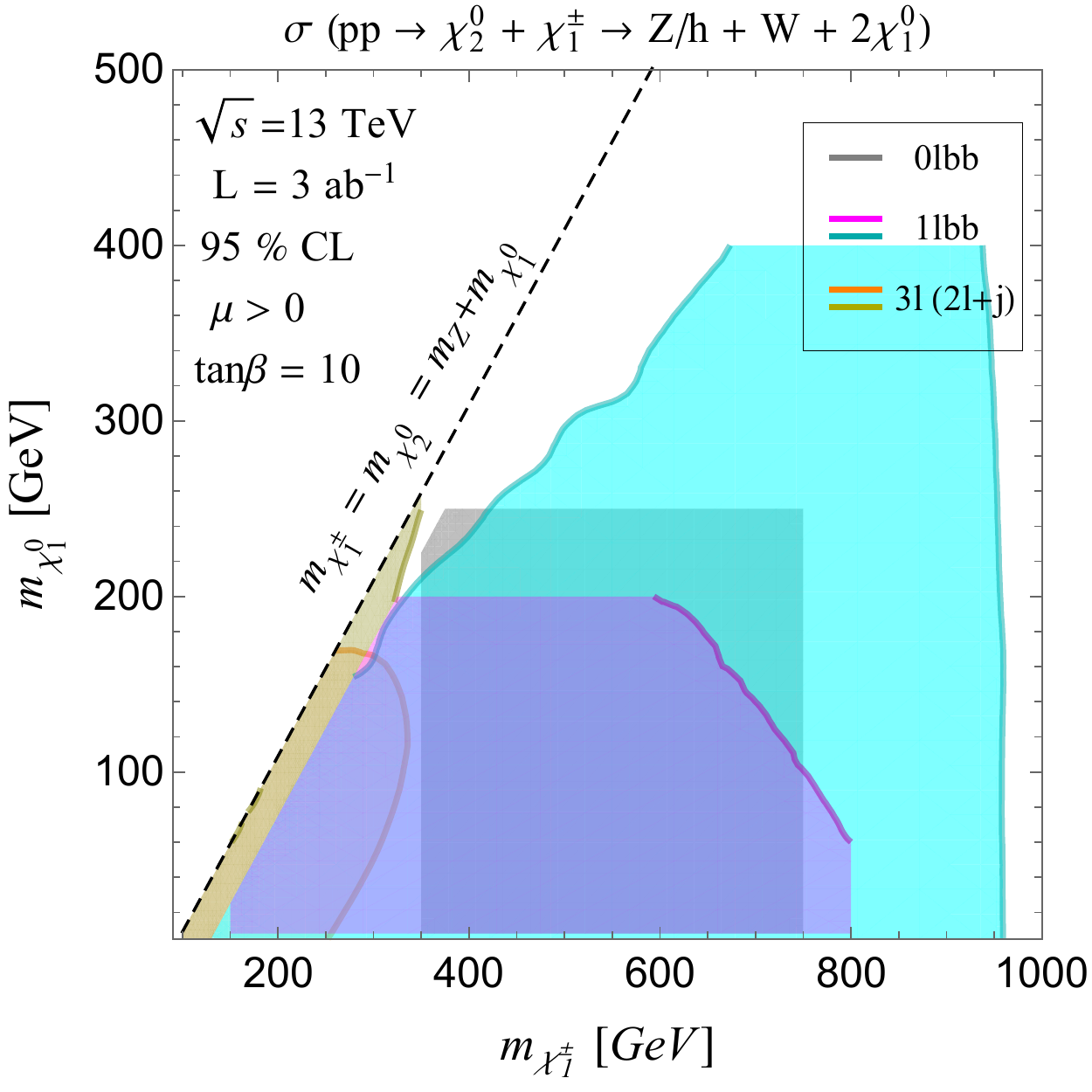} 
	\includegraphics[width=0.43 \columnwidth]{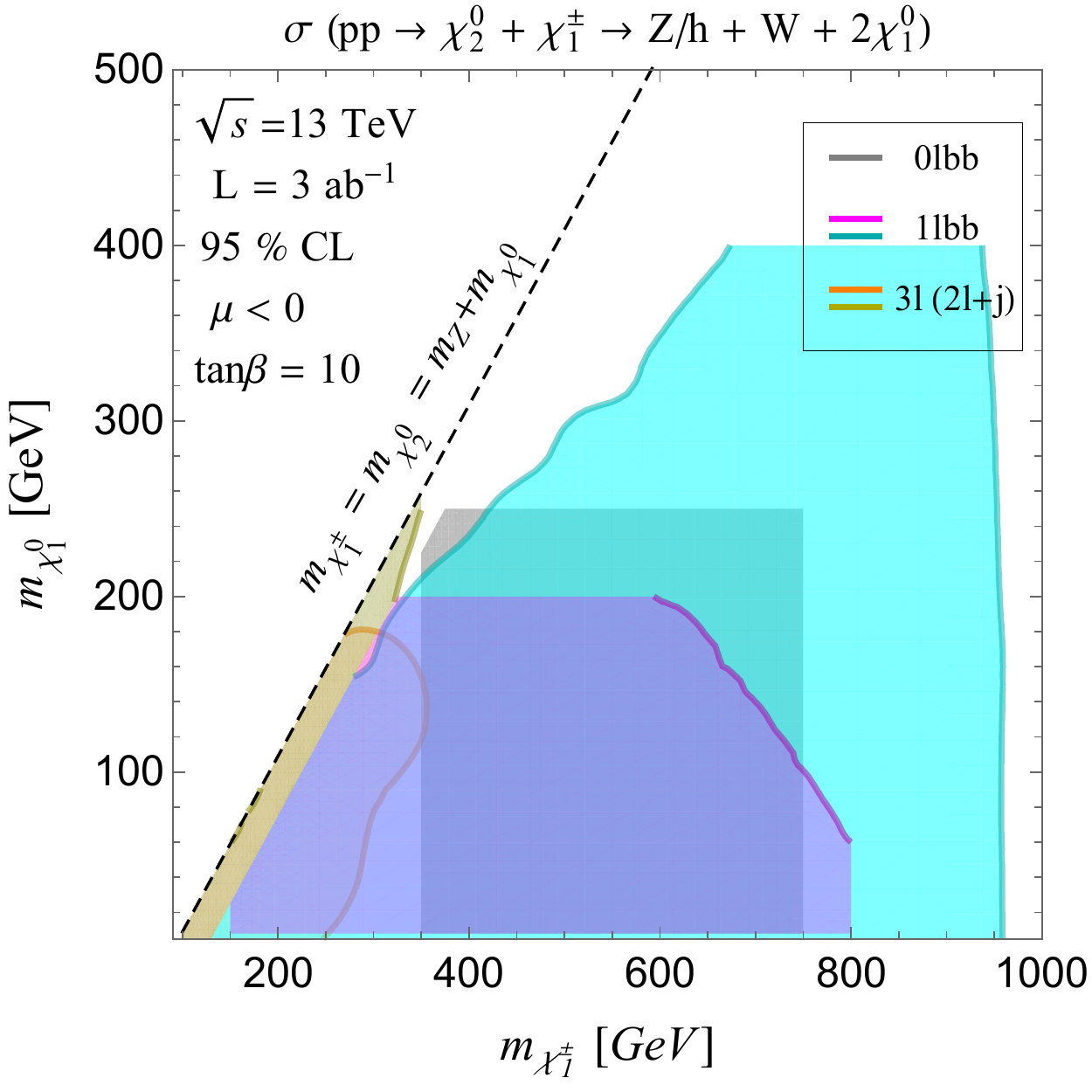} \\
	\includegraphics[width=0.43\columnwidth]{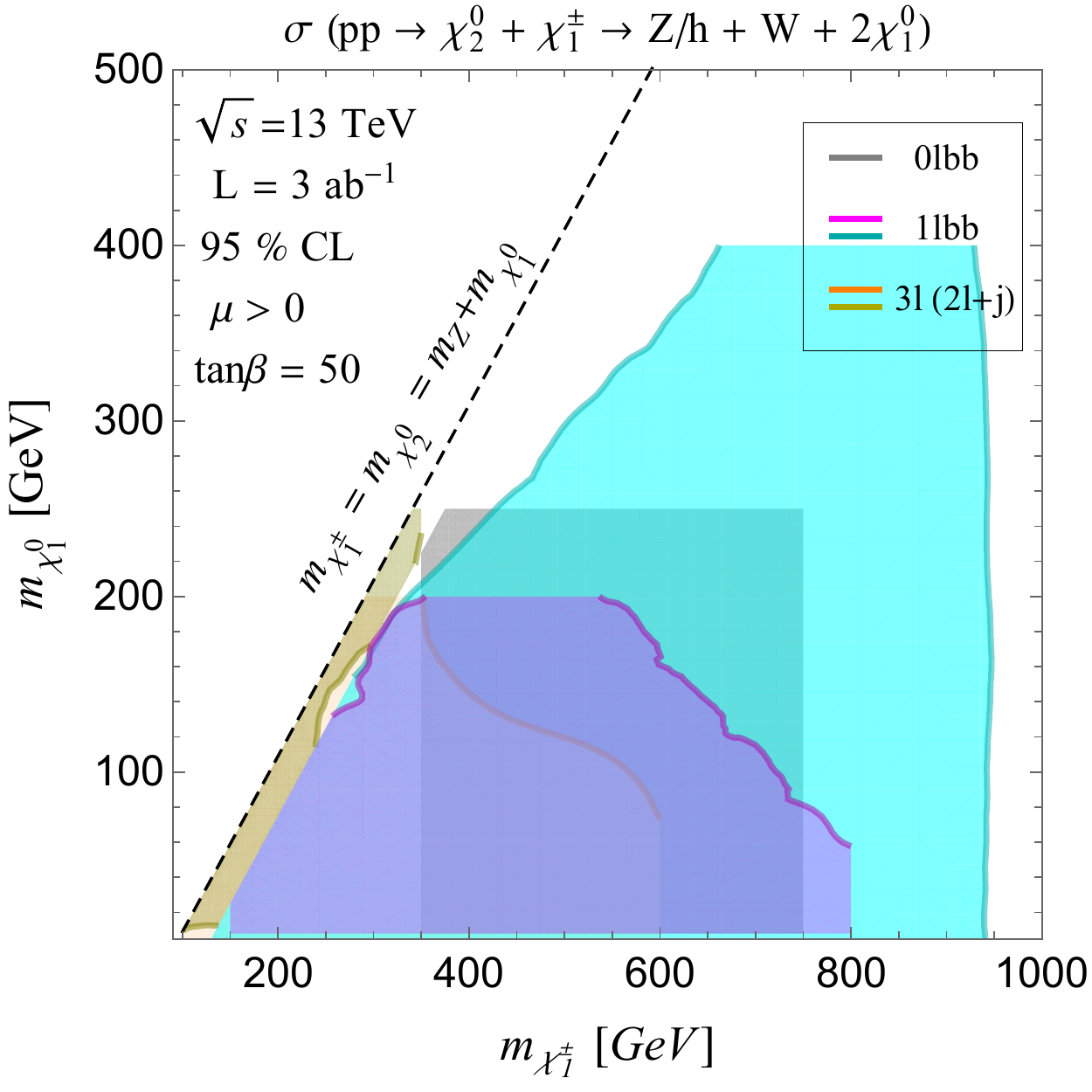} 
	\includegraphics[width=0.43 \columnwidth]{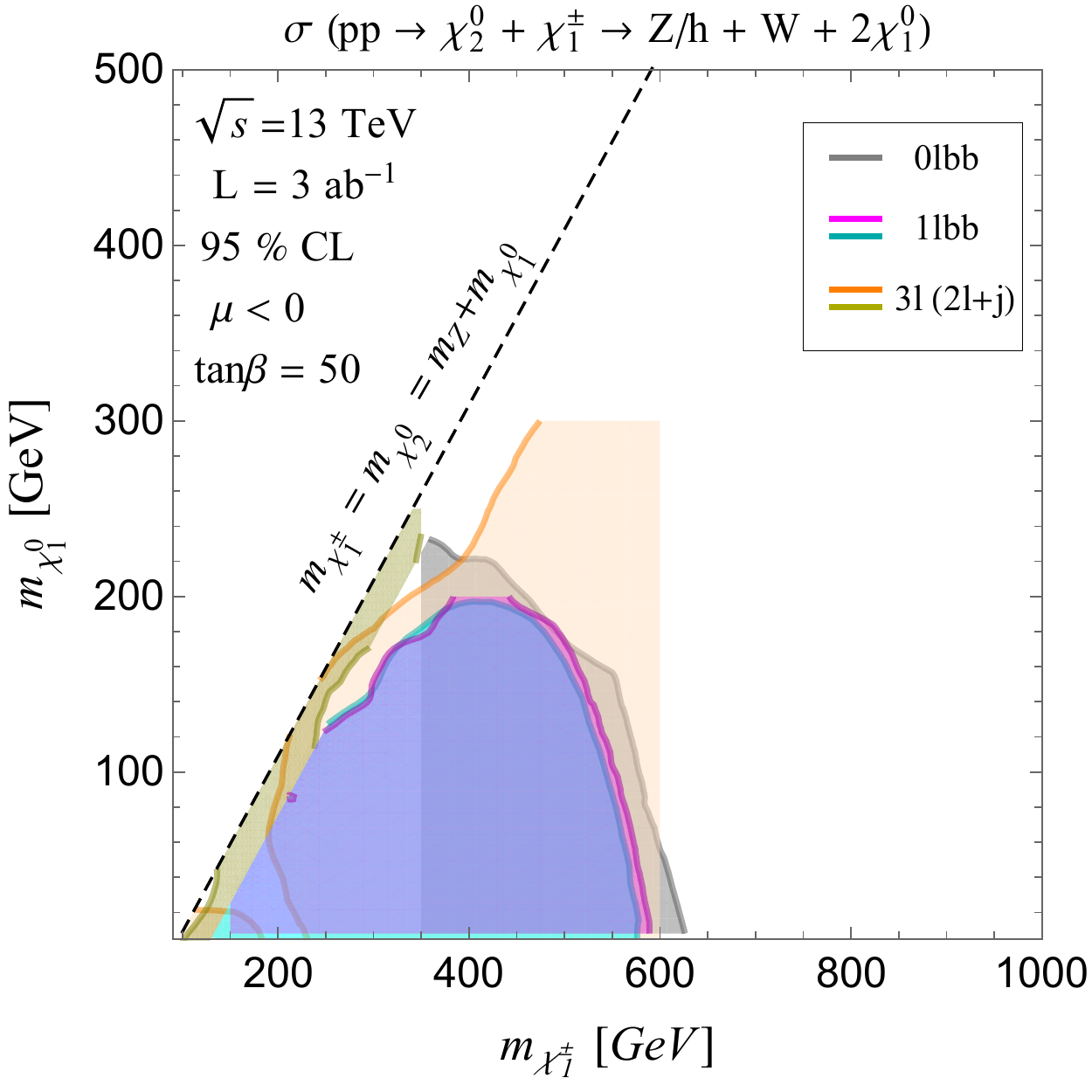} 
\caption{ Constraints on the Bino-Wino scenario projected at 3 ab$^{-1}$ at $95\%$ confidence level with $M_{SUSY}=|\mu|=10$ TeV. The labels are similar to Fig.~\ref{fig:2TeV_constraints}.}
\label{fig:10TeV_constraints-3ab}
\end{figure}
\begin{figure}[h!]
\centering
	\includegraphics[width=0.43 \columnwidth]{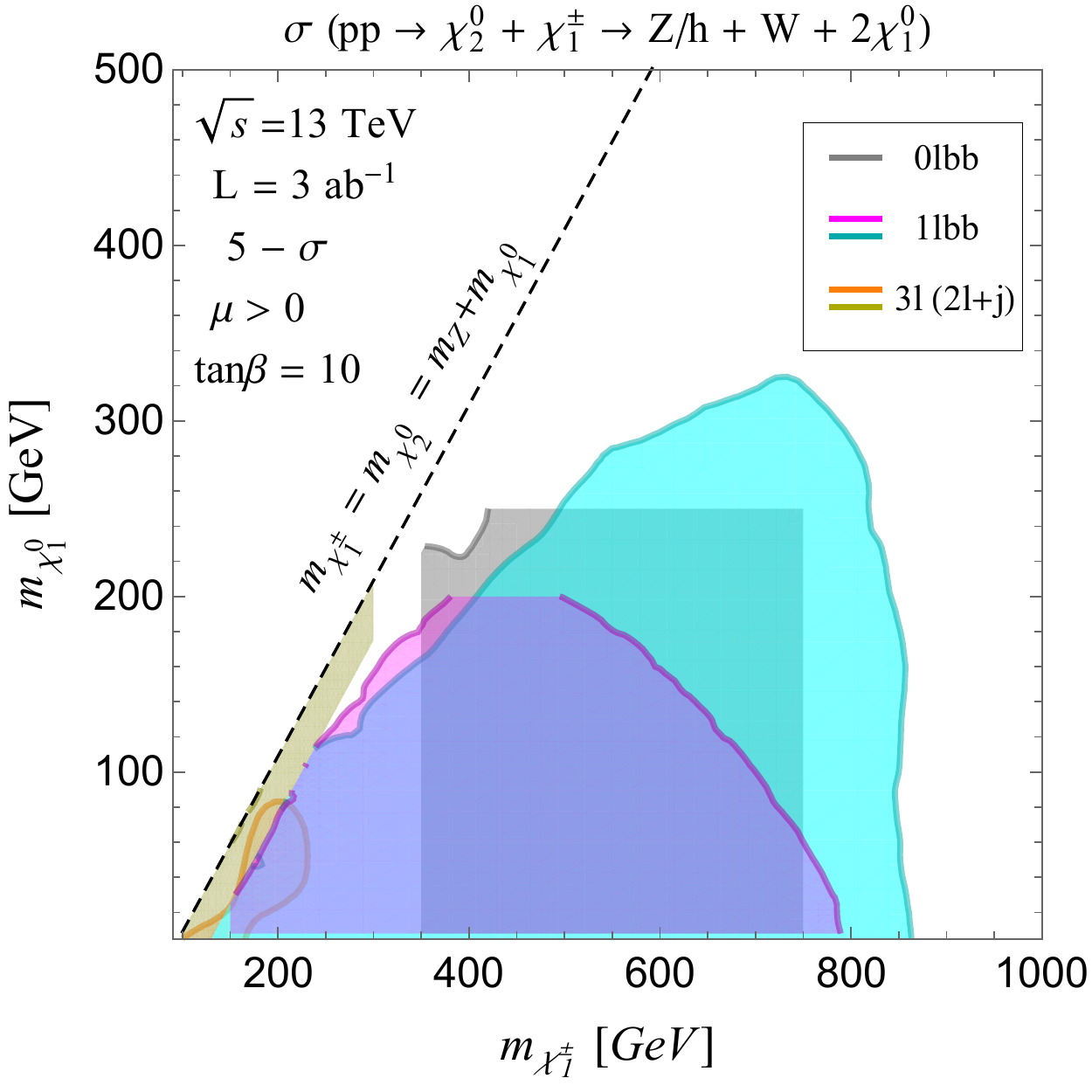} 
	\includegraphics[width=0.43 \columnwidth]{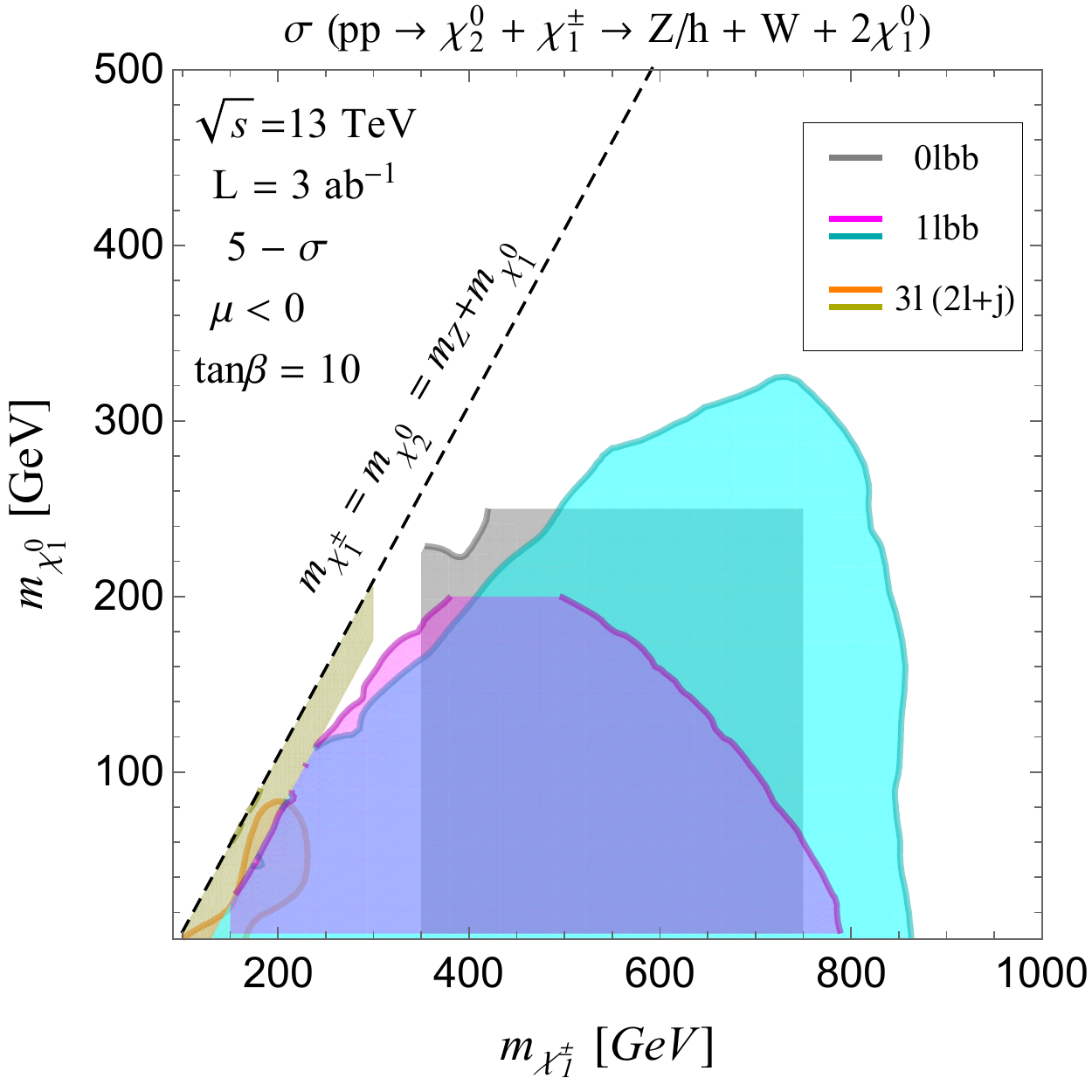} \\
	\includegraphics[width=0.43 \columnwidth]{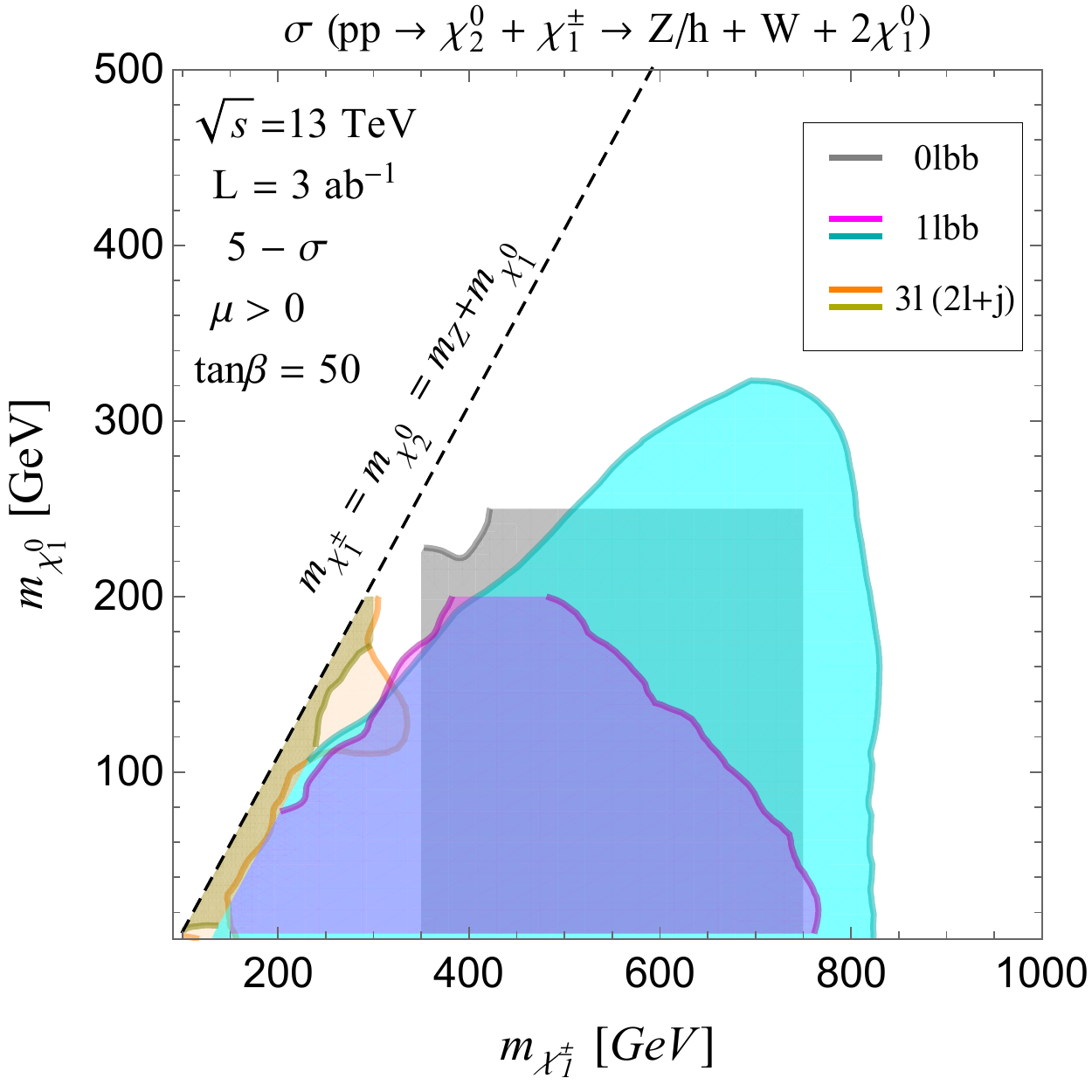} 
	\includegraphics[width=0.43 \columnwidth]{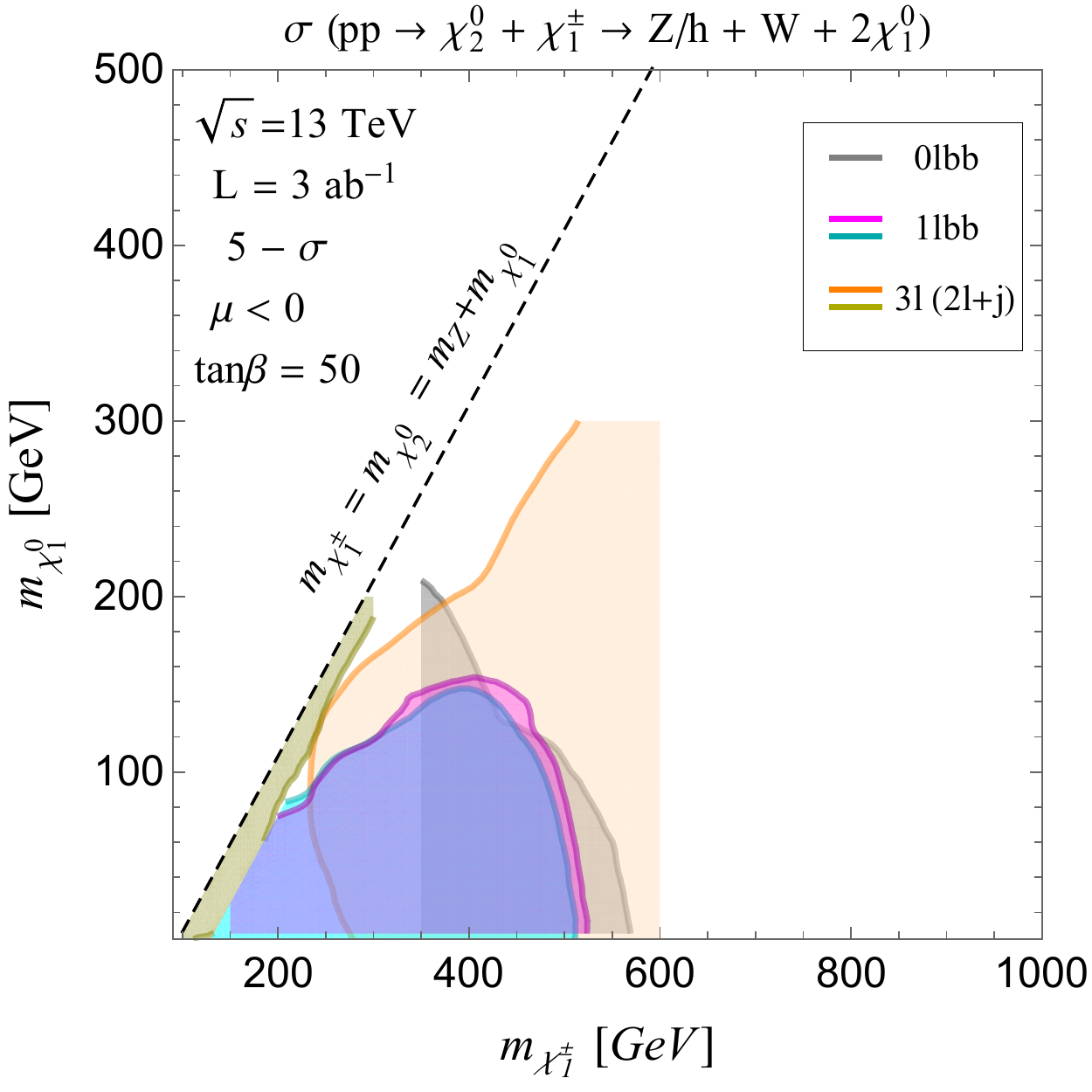} 
\caption{The future potential on the Bino-Wino scenario projected for $5~\sigma$ reach at 3 ab$^{-1}$ with $M_{SUSY}=|\mu|=10$ TeV. The labels are similar to Fig.~\ref{fig:2TeV_constraints}.}
\label{fig:10TeV_constraints-5sigma}
\end{figure}

 
When $M_{SUSY}=10$ TeV the production cross section reaches maximal values over the whole range of masses giving the strongest expected reach at the HL-LHC. In Fig.~\ref{fig:10TeV_constraints-3ab} and Fig.~\ref{fig:10TeV_constraints-5sigma}, we show the corresponding $95\%$ CL and $5-\sigma$ discovery bounds with $M_{SUSY}=|\mu|$. Here the $95\%$ CL bounds on chargino masses from Higgs decay searches begin to reach the TeV scale and beyond 400 GeV for neutralinos. In this case, the discovery region extends to $m_{\chi_1^\pm} \gtrsim 750$ GeV and $m_{\chi_1^0} \gtrsim 250$ GeV, and is significantly stronger than the current bounds, shown in Fig.~\ref{fig:10TeV_constraints}, implying again a strong discovery potential at the HL-LHC run.

\section{Conclusions}
\label{sec:conclusion}

The search for electroweak interacting particles is, together with precision measurements of the Higgs couplings, one of the most
promising activities in the HL-LHC era. In this article, we critically reanalyzed the search for electroweakinos in the
case of a Higgsino mass parameter significantly larger than the weak scale.  We showed that the signatures of Wino production
depend crucially on three parameters :  The first and second generation squark masses, which control the t-channel contribution
to the Wino production cross section, the sign of $\mu$, which control the mixing parameter determining the decay of the neutral
Winos into $Z$ or $h$ final states, and finally the relative size of the ratio of the Higgsino mass parameter to the average gaugino mass
to $\tan\beta$, which control the proximity to the blind spot for the decay of the neutral Wino into Higgs states for negative values of~$\mu$.

The t-channel contribution to the cross section interferes destructively with the s-channel one and hence the cross section becomes
larger for larger squark masses.  This destructive interference is still sizable for squark masses of the order of 2~TeV, but becomes
weak for squark masses above the 5~TeV scale, for which the maximal reach is  therefore achieved.  These very large values of the squark
masses are implicitly assumed in the experimental presentation of the LHC bounds for Wino-like particles decaying into
lighter Bino states.  It is important to stress that such dependence is not present in the production of Higgsino states, which couple
with the first and second generation squarks via their small Yukawa couplings. We refer to Ref.~\cite{Liu:2020muv} for the 
Higgsino search analysis. 

In general, the Higgs decay mode provides the dominant decay branching ratio of the neutral Winos and hence the tri-lepton channel
looses significance unless the mass difference between the neutral Winos and Binos are below the Higgs mass scale or
one is in the proximity of the previously mentioned blind spot. For positive values of $\mu$ with respect to the average
gaugino masses and large mass differences, the Branching ratio of the Higgs decay is very close to one. The blind spot
only occurs for negative values of $\mu$, in which case there may be a rich interplay between the Higgs decay and $Z$ decay
searches.

Two relevant conclusions of this study is that, depending on the parameters, the current exclusions limits may be
significantly weaker than the ones displayed in experimental searches and, most importantly, the discovery reach
of the HL-LHC greatly exceeds the region probed at current luminosities. This, together with 
similar results obtained in the case of Higgsino searches~\cite{Liu:2020muv},  provides a strong
motivation for the future electroweakino searches in the high luminosity LHC era.

\section*{Acknowledgments}
Work at University of Chicago is supported in part by U.S. Department of Energy grant
number DE-FG02-13ER41958. Work at ANL is supported in part by the U.S. Department
of Energy under Contract No. DE-AC02-06CH11357. NM acknowledges support by the U.S.
Department of Energy, Office of Science, Office of Work- force Development for Teachers and
Scientists, Office of Science Graduate Student Research (SCGSR) program. The SCGSR
program is administered by the Oak Ridge Institute for Science and Education (ORISE) for
the DOE. ORISE is managed by ORAU under contract number de-sc0014664.
JL acknowledges support by Peking University under Special Research grant number 7101502458.

\bibliography{ref}

\bibliographystyle{JHEP}

\end{document}